\documentclass[fleqn,usenatbib,useAMS]{mnras}
\usepackage{newtxtext,newtxmath}
\usepackage[T1]{fontenc}
\usepackage{ae,aecompl}

\usepackage{graphicx}   
\usepackage{amsmath}    
\usepackage{amssymb}    
\usepackage{pdflscape}  

\title[3D CCSN Explosion]{A Successful 3D Core-Collapse Supernova Explosion Model}

\author[D. Vartanyan et. al]{David Vartanyan$^{1}$\thanks{E-mail: dvartany@princeton.edu},
Adam Burrows$^{1}$,
David Radice$^{1,2}$ ,
M. Aaron Skinner$^{3}$,
\newauthor
Joshua Dolence$^{4}$
\\
$^{1}$Department of Astrophysical Sciences, Princeton University, Princeton, NJ 08544\\
$^{2}$ Institute for Advanced Study, 1 Einstein Dr, Princeton NJ 08540\\
$^{3}$Lawrence Livermore National Laboratory, P.O. Box 808, Livermore, CA 94551-0808\\
$^{4}$CCS-2, Los Alamos National Laboratory, P.O. Box 1663 
Los Alamos, NM 87545\\
}

\begin{document}
\label{firstpage}
\pagerange{\pageref{firstpage}--\pageref{lastpage}}
\maketitle

\begin{abstract}

In this paper, we present the results of our three-dimensional, multi-group, multi-neutrino-species radiation/hydrodynamic simulation using the state-of-the-art code F{\sc{ornax}} of the terminal dynamics of the core of a non-rotating 16-M$_{\odot}$ stellar progenitor.  The calculation incorporates redistribution by inelastic scattering, a correction for the effect of many-body interactions on the neutrino-nucleon scattering rates, approximate general relativity (including the effects of gravitational redshifts), velocity-dependent frequency advection, and an implementation of initial perturbations in the progenitor core.  The model explodes within $\sim$100 milliseconds of bounce (near when the silicon-oxygen interface is accreted through the temporarily-stalled shock) and by the end of the simulation (here, $\sim$677 milliseconds after bounce) is accumulating explosion energy at a rate of $\sim$2.5$\times$10$^{50}$ erg s$^{-1}$. The supernova explodes with an asymmetrical multi-plume structure, with one hemisphere predominating.  The gravitational mass of the residual proto-neutron star at $\sim$677 milliseconds is $\sim$1.42 M$_{\odot}$. Even at the end of the simulation, explosion in most of the solid angle is accompanied by some accretion in an annular region at the wasp-like waist of the debris field. The ejecta electron fraction (Y$_e$) is distributed between $\sim$0.48 and $\sim$0.56, with most of the ejecta mass proton-rich.  This may have implications for supernova nucleosynthesis, and could have a bearing on the p- and $\nu$p-processes and on the site of the first peak of the r-process. The ejecta spatial distributions of both Y$_e$ and mass density are predominantly in wide-angle plumes and large-scale structures, but are nevertheless quite patchy.

\end{abstract}

\section{Introduction}
\label{sec:1}

The neutrino mechanism of core-collapse supernovae (CCSNe) was proposed  
more than fifty years ago (\citealt{1966ApJ...143..626C}), but due to the
complexity and exotic character of the environment in which it occurs and
the realization that hydrodynamic instabilities and turbulence are crucial
to explosion in all but a small subset of progenitor stars, credible
confirmation of this mechanism and its observational validation have been
frustratingly slow. Along with the requirement to incorporate nuclear and 
particle physics that does justice to the wide range of relevant
neutrino-matter interactions and to the equation of state of dense nuclear
matter, the centrality of turbulent convective and shock instabilities
that break spherical symmetry has necessitated performing theoretical     
simulations in multiple spatial dimensions.  The two-dimensional (2D)
simulations (axisymmetric) of the 1990's lacked detailed neutrino physics,
but demonstrated the relevance of neutrino-driven convection 
(\citealt{herant1994,bhf1995}). The early years of this   
millenium introduced another instability (the standing-accretion shock instability, \citealt{blondin2003}) and
subsequent work built on previous 2D efforts by incorporating general
relativity (GR, at various levels of approximation), improving the
physical fidelity of the neutrino interactions embedded into the codes,   
enhancing the spatial resolution of the calculations, and carrying
simulations out to later physical times. Summaries of some of this history
can be found in reviews by \cite{janka2012}, \cite{burrows2013}, \cite{muller2016}, and \cite{jms2016}.
In fact, progress in understanding the CCSN mechanism has   
paralleled progress in both physics and computational capability, and such
progress has spanned decades.  It is only recently that fully
three-dimensional (3D) simulations with multi-group neutrino transfer that
address all the physical terms and effects, employ state-of-the-art
nuclear equations-of-state, and calculate for a physically significant
duration have emerged. Though there is still much work to do, the recent 
advent of codes that address the full dimensional and physics requirements
of the CCSN problem represents a watershed in the theoretical exploration 
of the supernova mechanism. In this paper, we present one such modern
simulation of the explosion in 3D of a 16-M$_{\odot}$ star, using our new supernova code F{\sc{ornax}} (\citealt{skinner2016, 2017ApJ...850...43R,vartanyan2018a,burrows2018,2018arXiv180607390S}).

State-of-the-art calculations in 3D exploring the mechanism of CCSN
explosions have undergone significant evolution and improvement over the
years. Sixteen years ago, \cite{2002ApJ...574L..65F,2004ApJ...601..391F}
used a smooth-particle hydrodynamics code SNSPH to explore the differences
between 2D and 3D simulations and the possible role of rapid rotation.
They found their 2D and 3D simulations were similar and that rapid rotation modified
the driving core neutrino emissions. However, these simulations employed gray radiation,
did not include inelastic energy redistribution nor velocity dependent
transport effects, and ignored GR effects. Parameterized studies in 3D
(\citealt{2010ApJ...720..694N, 2013ApJ...765..110D,
2012ApJ...755..138H,2014ApJ...785..123C, 2015ApJ...799....5C}) disagreed on
the relative difficulty of explosion in 3D vs. 2D. However, these
simulations, while boasting improved hydrodynamics algorithms and
resolution, used ``lightbulb" neutrino driving and did not employ
competitive neutrino transfer and microphysics. Using ZEUS-MP on a
low-resolution 3D grid, \cite{2012ApJ...749...98T} witnessed the explosion
of a 11.2-M$_{\odot}$ progenitor (\citealt{2002RvMP...74.1015W}). However,
these authors used the sub-optimal IDSA scheme neutrino transport approach
(\citealt{idsa}), which ignores velocity-dependence, GR, and
inelasticity, stitches together the opaque and transparent realms in an ad
hoc fashion, uses the problematic ``ray-by-ray" approach to multi-D
transport, and either neglects ``heavy" neutrinos or incorporates them in
a ``leakage" format. The ray-by-ray approach used by many early and
current studies performs multiple one-dimension transport calculations, in
lieu of truly multi-D transport, and thereby ignores the important effects
of lateral transport (\citealt{skinner2016}).

Using the CHIMERA code, \cite{2015ApJ...807L..31L} witnessed the explosion
of a 15-M$_{\odot}$ progenitor star (\citealt{wh07})
$\sim$300 milliseconds (ms) after bounce, $\sim$100 ms later than their 2D
simulation. These authors used state-of-the-art microphysics and
approximate GR, but used multi-group flux-limited diffusion and the
reduced-dimension ray-by-ray approach and evolved the inner 6-8 kilometers (km) in
spherical symmetry. In addition, they employed the LS220 equation of state
(EOS) (\citealt{1991NuPhA.535..331L}), now known to be inconsistent with  
known nuclear systematics. 

Early low-resolution 3D simulations using the Prometheus-Vertex code
(\citealt{2013ApJ...770...66H, 2014ApJ...792...96T}) found that the
11.2-, 27- (\citealt{2002RvMP...74.1015W}), and
25-M$_{\odot}$ (\citealt{wh07}) non-rotating progenitors
did not explode in 3D, while their 2D counterparts did.
Prometheus-Vertex uses state-of-the-art neutrino microphysics, a
multi-group variable Eddington factor transport algorithm with approximate
GR (\citealt{marek2006}), but uses the ray-by-ray+ approximation to neutrino
transport. Later, this group \citep{2015ApJ...801L..24M} witnessed the
explosion of a zero-metallicity 9.6-M$_{\odot}$ progenitor in 3D, a model 
that explodes easily in 1D (\citealt{2017ApJ...850...43R}). By making a large
strangeness correction to the axial-vector coupling constant in
Prometheus-Vertex, \cite{2015ApJ...808L..42M} were able to generate an    
explosion in 3D of the non-rotating 20-M$_{\odot}$ progenitor that did not
otherwise explode. However, such a large correction may be inconsistent
with nuclear experiment (\citealt{ahmed2012,green2017}). Recently,
this group (\citealt{2018ApJ...852...28S}) has found that rapidly rotating
progenitor models (\citealt{2005ApJ...626..350H}) explode shortly after
the accretion of the silicon-oxygen (Si-O) interface. They argue, as do \cite{takiwaki2016}, 
that a strong non-axisymmetric spiral mode facilitates explosion in the 
rapidly-rotating context. However, with their
default neutrino physics, this group has yet to witness the explosion in 3D of any non-rotating models using Prometheus-Vertex. 
Moreover, their 3D models were all calculated using the LS220 EOS.


Using the Coconut-FMT code in 3D, \cite{2015MNRAS.453..287M} witnessed the
explosion of the 11.2-M$_{\odot}$ progenitor of \cite{2002RvMP...74.1015W}
in 3D employing the LS220 nuclear EOS. However, Coconut-FMT employs 
simplified multi-group neutrino transport, the ray-by-ray approximation,
neglects both velocity dependence in the neutrino sector and inelastic
scattering, and cuts out the proto-neutron star (PNS) core.  Its virtue is
that it incorporates conformally-flat GR. Using Coconut-FMT and a 3D  
18-M$_{\odot}$ initial progenitor to provide perturbations,
\cite{muller2017} witnessed what they interpret as a
perturbation-aided explosion and the simulation was carried out to an
impressive $\sim$2.5 seconds after bounce.

It is only recently that codes with truly multi-dimensional, multi-group
transport, without the ray-by-ray compromise and with state-of-the-art
microphysics and approximate or accurate GR, have been constructed and
fielded. \cite{2016ApJ...831...98R} used the adaptive-mesh-refinement     
(AMR) Cartesian code Zelmani with full GR, the M1 moment closure approach,
the SFHo nuclear EOS (\citealt{2013ApJ...774...17S}), but without velocity
dependence in the transport sector or inelastic scattering, to evolve the
27-M$_{\odot}$ progenitor of \cite{2002RvMP...74.1015W}. Using the same   
code, \cite{ott2018_rel} explored the 12-, 15-, 20-, 27-, and
40-M$_{\odot}$ progenitor models of \cite{wh07}. This   
team witnessed the low-energy explosion of all models, save the
12-M$_{\odot}$ model. More recently, \cite{2016ApJS..222...20K}
have developed a multi-group radiation-hydrodynamic CCSN code with M1
closure, detailed microphysics, and full GR. However, their recent CCSN
simulations (\citealt{2016ApJ...829L..14K}) of 11.2-, 40-
(\citealt{2002RvMP...74.1015W}), and 15-M$_{\odot}$
(\citealt{1995ApJS..101..181W}) progenitors were done with gray
transport and none of their models exploded. A related code using the
FLASH architecture, AMR, state-of-the-art microphysics, approximate GR,
and M1 transport more recently witnesses no explosion for a 20-M$_{\odot}$ progenitor, but noted large asymmetries in the Si and O shells that might dynamically aid explosion (\citealt{oconnor_couch2018}).

In this paper, we present the first results in a series of 3D simulations
that employ our new code F{\sc{ornax}} (\citealt{2018arXiv180607390S}).
F{\sc{ornax}} is a multi-group, velocity-dependent neutrino transport code
that employs the M1 two-moment closure scheme. It incorporates
state-of-the-art neutrino microphysics, approximate GR (with gravitational
redshifts), inelastic energy redistribution via scattering, and does not  
employ the ray-by-ray simplification. Furthermore, it uses a dendritic
grid that deresolves in angle upon approach to the core, while maintaining
good zone sizes. This allows us to include the stellar center while
employing a spherical grid but without incurring an onerous Courant time step
penalty.  We find that the 16-M$_{\odot}$ progenitor
(\citealt{wh07}) explodes in 3D, and does so shortly before
its 2D counterpart.

Throughout this paper, we explore the dimensional dependence (2D vs. 3D) of the explosion properties. We organize the paper as follows: In \S\ref{sec:outline}, we outline the setup of our simulation. We explore the basic explosion properties in the beginning of \S\ref{sec:expl} and the shock evolution in \S\ref{sec:shock}. In \S\ref{sec:energy} and \ref{sec:lum}, we explore the explosion energetics and heating rates and the luminosities and mean energies, respectively. We look at the ejecta composition in \S\ref{sec:ejecta}, and study PNS convection in \S\ref{sec:pns}. We comment in \S\ref{sec:sasi} on the possibility of the lepton-number emission self-sustained asymmetry (LESA; \citealt{2014ApJ...792...96T}) in our 3D simulation and the lack of the standing accretion-shock instability (SASI). In \S\ref{sec:conc}, we conclude with summary comments and observations.

\section{Numerical Setup and Methods}
\label{sec:outline}

The progenitor upon which we focus in this paper is the 16-M$_{\odot}$ 
model of \cite{wh07} (which was studied in 2D in \citealt{vartanyan2018a}), and we employ
the state-of-the-art multi-D, multi-group radiation/hydrodynamic code F{\sc{ornax}} (Skinner et al. 2018).
Earlier supernova work using F{\sc{ornax}} includes  \citealt{2016ApJ...817..182W} (neutrino breakout burst 
detection), \citealt{skinner2016} (shortcomings of the ray-by-ray approximation in core-collapse simulations), 
\citealt{2017ApJ...850...43R} (low-mass CCSNe),\, \citealt{burrows2018} (the role 
of microphysics in CCSNe), \citealt{2018arXiv180101914M} (gravitational wave signatures of CCSNe), 
\citealt{vartanyan2018a} (CCSNe from 12-25 M$_{\odot}$), and \citealt{shaquann2018} 
(neutrino detection of CCSNe). 

F{\sc{ornax}} is a multi-dimensional, multi-group radiation hydrodynamics code originally
constructed to study core-collapse supernovae and its structure, capabilities, and variety of code tests are 
described in \cite{2018arXiv180607390S}. The generalization of the equations to approximate
general-relativistic gravity and redshifts is described in Appendix \ref{redshift}. 
In 2D and 3D, F{\sc{ornax}} employs a dendritic grid which deresolves at small radii and in 3D along the $\phi$ axis 
to avoid overly-restrictive CFL time step limitations, while at the same time preserving cell size and 
aspect ratios.  Our method of deresolving near the polar axis for 3D simulations allows us to partially
overcome axial artifacts seen conventionally in 3D simulations in spherical coordinates 
(see, e.g. \citealt{2015ApJ...807L..31L,muller2017}). F{\sc{ornax}} solves the 
comoving-frame velocity-dependent transport equations to order O($v/c$). The hydrodynamics 
uses a directionally-unsplit Godunov-type finite-volume scheme and computes fluxes at cell 
interfaces using an HLLC Riemann solver. For the 3D simulation highlighted in this paper, we employ a spherical grid in r, $\theta$, and $\phi$
of resolution 608$\times$128$\times$256.  For the comparison 2D simulation, the axisymmetric grid 
has resolution 608$\times$128.  
The radial grid extends out to 10,000 kilometers (km) and is spaced evenly with $\Delta{r}\sim0.5$ km 
for r $\lesssim$ 50 km and logarithmically for r $\gtrsim$ 50 km, with a smooth transition in between. 
The angular grid resolution varies smoothly from $\Delta\theta \sim$1.9$^\circ$ at the poles 
to $\Delta\theta \sim$1.3$^\circ$ at the equator, and has $\Delta\phi \sim$1.4$^\circ$ uniformly. For this project, we used a monopole approximation 
for relativistic gravity following \cite{marek2006}, as described in Appendix \ref{redshift}, 
and employed the SFHo equation of state (\citealt{2013ApJ...774...17S}), which is 
consistent with all currently known nuclear constraints (\citealt{2017ApJ...848..105T}). 

We solve for radiation transfer using the M1 closure scheme for the second and 
third moments of the radiation fields (\citealt{2011JQSRT.112.1323V}) and follow 
three species of neutrinos: electron-type ($\nu_{e}$), anti-electron-type ($\bar{\nu}_{e}$), 
and ``$\nu_{\mu}$"-type ($\nu_{\mu}$, $\bar{\nu}_{\mu}$, $\nu_{\tau}$, and $\bar{\nu}_{\tau}$ 
neutrino species collectively). We use 12 energy groups spaced logarithmically between 1 and 300 MeV 
for the electron neutrinos and to 100 MeV for the anti-electron- and ``$\nu_{\mu}$"-neutrinos.
The neutrino-matter interactions implemented and the handling of 
inelastic scattering and energy redistribution of neutrinos 
off electrons and nucleons are both summarized in Appendix \ref{sec.nu_inter}.

For this paper, we initially evolve collapse in 1D until 10 ms after bounce, and 
then map to higher dimensions. After mapping, we impose velocity perturbations (for the 2D and 3D, but not 1D, simulations) following \cite{muller_janka_pert}
in three spatially distinct regions (50 $-$ 85 km, 90 $-$ 250 km, and 260 $-$ 500 km),  
with a maximum speed of 500 km s$^{-1}$ and harmonic quantum numbers of $l$ = 2, $m$ = 1, 
and $n$ = 5 (radial), as defined in \cite{muller_janka_pert}. These perturbations were motivated by 
\cite{2016ApJ...833..124M}, who evolve the last minutes of a 3D progenitor and find convective velocities of almost 1000 km s$^{-1}$ at the onset of collapse (with a corresponding Mach number of 0.1) in the O-shell around 5000 km with a prominent $l=2$ mode.

Our 3D simulation was evolved to 677 ms after core bounce, 
and required a total resource burn of $\sim$18 million CPU-hours on the NERSC/Cori II machine using 16256 cores in parallel. 
\footnote{For comparison, some earlier 3D simulations (e.g., \citealt{2018ApJ...852...28S} for a model with slightly lower resolution, but with rotation.)
required $\sim$50$-$100 million CPU-hours to evolve to 0.5 seconds after bounce.}

We note that the 16-M$_{\odot}$ progenitor was studied in \cite{vartanyan2018a}, but with a different setup. In that paper, we did not include initial velocity perturbations, had 20 (instead of 12) energy bins per neutrino species, and employed an angular resolution of 256 polar cells (instead of 128). We also did not map from 1D to 2D 10 ms after bounce, but evolved entirely in 2D. Our grid then extended to 20,000 km (not 10,000 km). For the 2D comparison model in this paper, we maintain these differences to mimic the setup we use for our concurrent 3D run. However, we obtain the same overall results for the 16-M$_{\odot}$ progenitor seen in \cite{vartanyan2018a}.

\section{Explosion Properties}
\label{sec:expl}

We find that the 16-M$_{\odot}$ progenitor of \cite{wh07} explodes in 
both the corresponding 2D and 3D simulations, at $\sim$100 and $\sim$120 ms after bounce, respectively. 
The corresponding 1D simulation does not explode. The explosions in 2D and 3D are abetted 
by the inclusion of detailed microphysics $-$ in particular, inelastic scattering off electrons,
nucleons, and the associated energy redistribution, and the decrease in the neutral-current neutrino-nucleon 
scattering rates due to the many-body effect (\citealt{burrows2018,PhysRevC.95.025801}) $-$ as well as a steep density gradient at the 
silicon-oxygen interface located deep within the progenitor, near an interior mass of 
$\sim$1.5-M$_{\odot}$ (\citealt{vartanyan2018a,ott2018_rel}). Unlike in many recent 3D simulations, 
we use the SFHo equation of state in this work.

At the end of our 3D simulation, $\sim$677 ms after bounce, the maximum shock radius 
has reached $\sim$5000 km, with an asymptotic velocity of $\sim$10,000 km s$^{-1}$. The diagnostic explosion 
energy is $\sim$1.7$\times$10$^{50}$ erg by this time. The mass of the core ejecta, defined as neutrino-processed 
gravitationally unbound material, is $\sim$0.08 M$_{\odot}$ and growing. The corresponding gravitational 
PNS mass is $\sim$1.42 M$_{\odot}$ and the mean PNS radius $\sim$29 km. These features are explored in greater detail 
in the later sections and compared to the results in 2D. In all regards, we find that integrated 3D metrics 
are significantly less variable with time than their 2D analogs.

In Fig.\,\ref{fig:1}, we show a time sequence of the entropy of the 3D simulation, illustrating 
the highly non-axisymmetric nature of the explosion. By $\sim$100 ms after bounce, shock expansion 
and explosion are underway, with the outflow initially constituting bubbles interior to the shock. The explosion  
assumes a bi-cameral structure, with the two hemispheres separated by a plane oriented with $\theta$ $\sim$40$^{\circ}$ and $\phi$ $\sim$50$^{\circ}$ in spherical coordinates. Unlike in 2D, the 
explosion does not orient around any coordinate axis and there is no axial sloshing; any explosion 
axis that emerges does so naturally and is not imposed.  Indeed, the explosion is not isotropic, but 
has a preferred direction, clockwise-orthogonal to the dividing plane. The left hemisphere (in this 
projection) dominates and we see some fingers along the axis at $\sim$443 ms (3rd panel), but these are accreted soon 
after. The electron fraction distribution follows the entropy distribution, with high Y$_e$ (Y$_e$ $>$ 0.53)
concentrated along the outer cusps of each plume (see \S\ref{sec:ejecta}). We see high-Y$_e$ material in both plumes 
as well as in the interior.

\begin{figure*}
\includegraphics[width=0.42\textwidth]{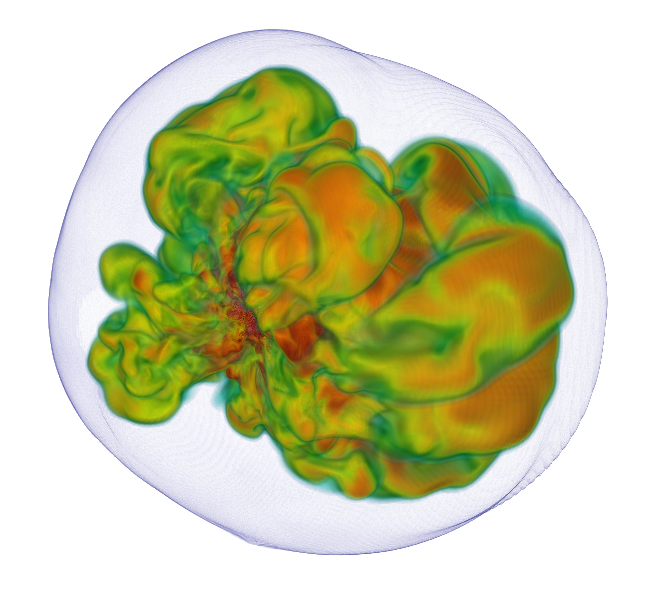}
\caption{Volume rendering of the entropy per baryon showing the morphology of the explosion 
of the 11-M$_{\odot}$ progenitor from \protect\cite{sukhbold2018}. The snapshot is 
taken at $\sim$690 ms after bounce, when the shock wave (blue outer surface in the figure) 
has an average radius of $\sim$3500 km. The shock is expanding quasi-spherically, however 
accretion continues on one side of the PNS, while neutrino-driven winds inflate higher-entropy bubbles on the other side.}
\label{fig:1_dradice}
\end{figure*}

We show in Fig.\,\ref{fig:1_dradice} a volume rendering of the entropy per baryon showing the morphology of the explosion of the 11-M$_{\odot}$ progenitor from \protect\cite{sukhbold2018} from an upcoming paper (Radice et. al, 2018). The snapshot is taken at $\sim$690 ms after bounce, when the shock wave (blue outer surface in the figure) has an average radius of $\sim$3500 km. The explosion behaves similarly to that of the 16-M$_{\odot}$ progenitor we evolve. The shock is expanding quasi-spherically; however, accretion continues on one side of the PNS, while neutrino-driven winds inflate high-entropy bubbles on the other side.

\begin{figure*}
\includegraphics[width=0.49\textwidth]{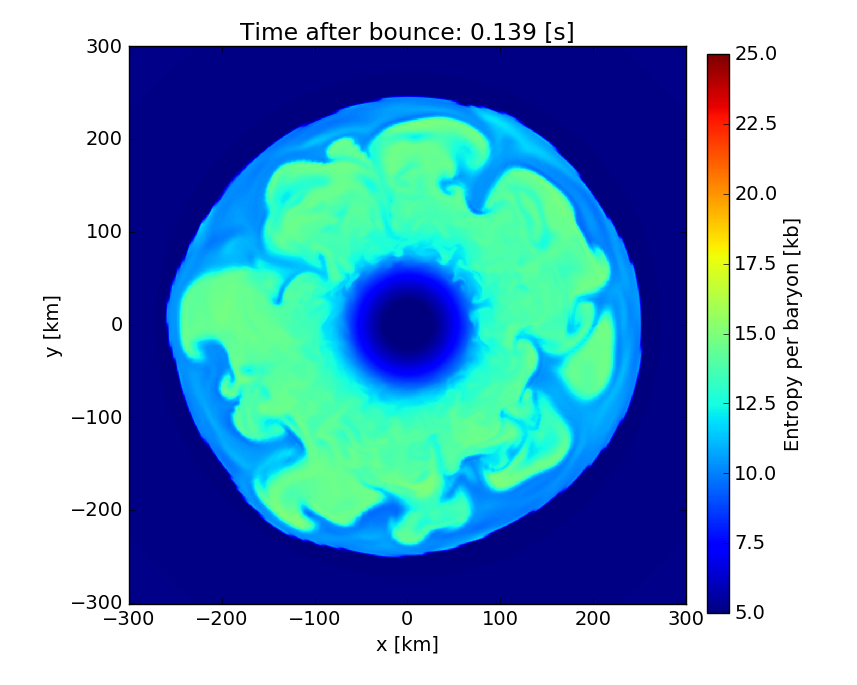}
\includegraphics[width=0.49\textwidth]{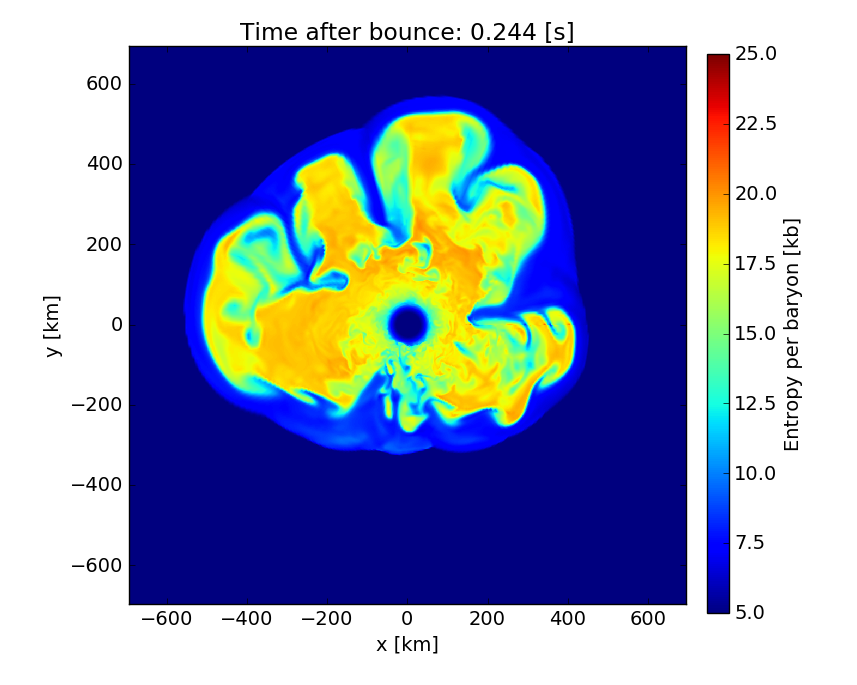}
\includegraphics[width=0.49\textwidth]{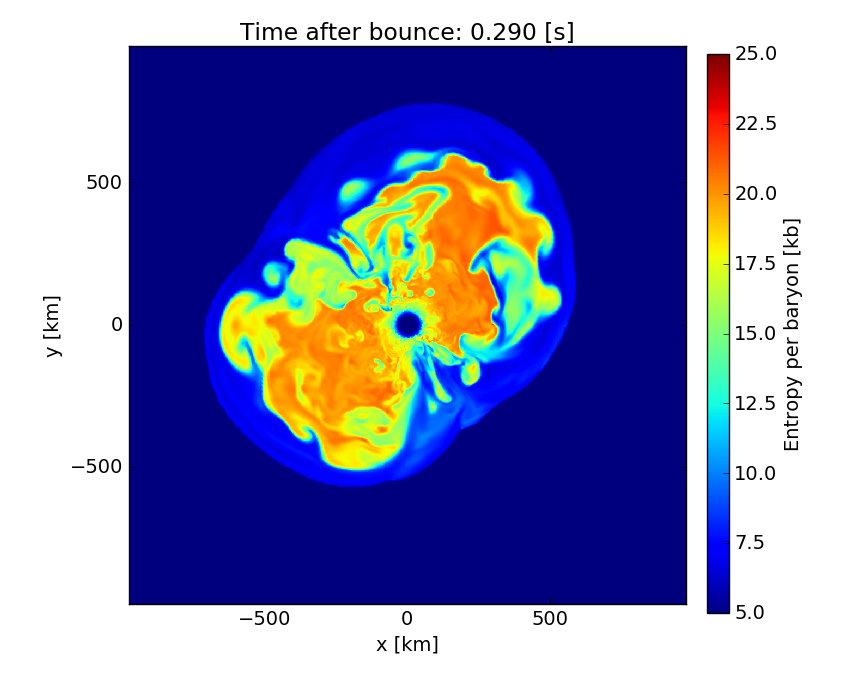}
\includegraphics[width=0.49\textwidth]{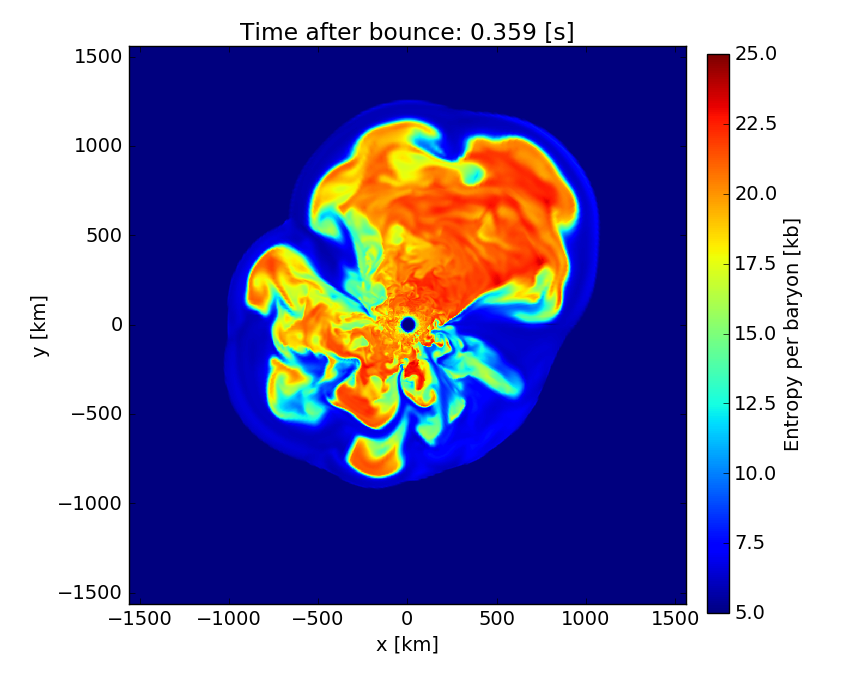}
\includegraphics[width=0.49\textwidth]{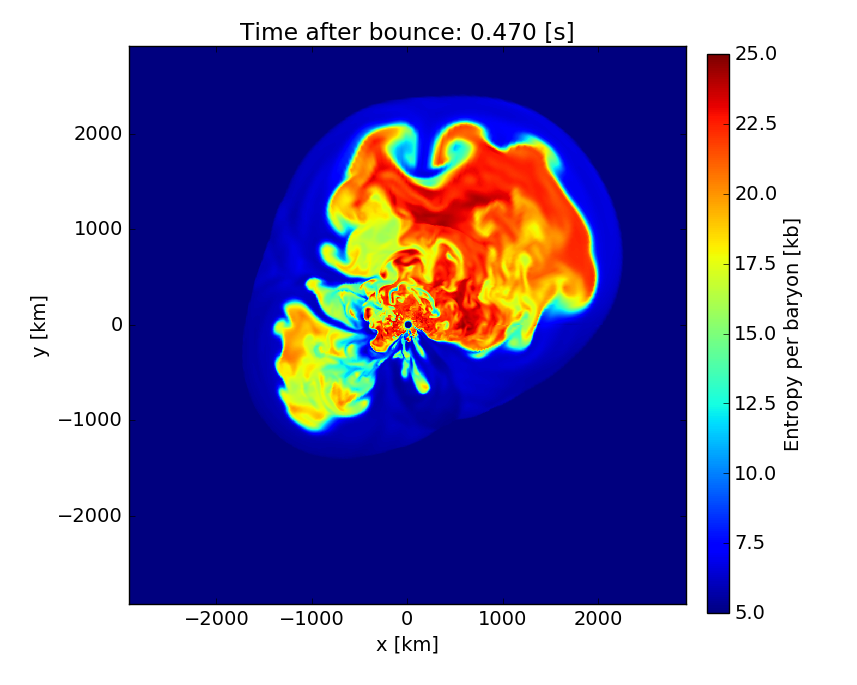}
\includegraphics[width=0.49\textwidth]{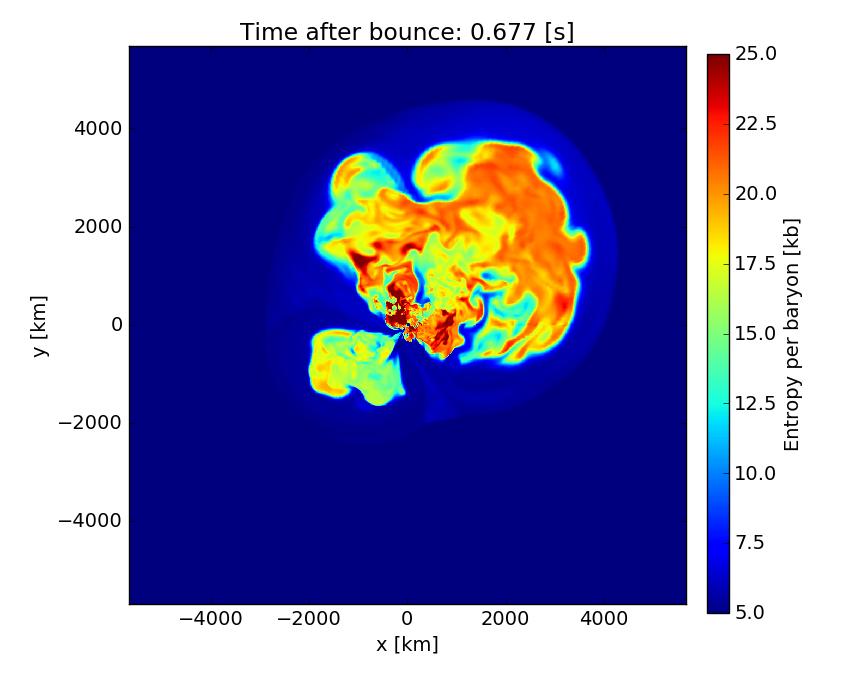}
\caption{Time sequence slices in the x-y plane illustrating the entropy of the 3D simulation of
the 16-M$_{\odot}$ progenitor. Note the changing spatial scales with time. At early times, shock expansion is
driven by multiple bubbles, which coalesce into larger plumes.  At approximately 300 ms after bounce,
we note the development of a dividing axis with two dominant plumes in this slicing. At late times,
a single dominant explosion plume emerges, seemingly at the expense of the secondary plume.
A persistent wind is present in both plumes initially, and finally, only in the dominant plume.
The secondary plume persists and grows, with a characteristic scale of $\sim$2000 km,
half the size of the primary plume at the end of our simulation. We see simultaneous explosion
and accretion. The shock evolution transitions from quasi-spherical expansion to axial expansion, with the axis
arbitrarily chosen. See the text for a discussion.}
\label{fig:2}
\end{figure*}

In Fig.\,\ref{fig:2}, we illustrate a time sequence of entropy slices for the 3D simulation of the 16-M$_{\odot}$ progenitor along the x-y plane (top). At early times, shock breakout is driven by multiple smaller bubbles in 3D, as opposed to a few large plumes in 2D. The shock evolution in 3D transitions from quasi-spherical expansion to expansion along an axis, with the axis randomly chosen. By $\sim$300 ms after bounce, the plumes in 3D have merged into two distinct larger-solid-angle bubbles oriented along a clear axis. We see matter cross and accrete through this axis at earlier times before the explosion settles into the final configuration (see panels 3-5 of Fig.\,\ref{fig:2}). At late times in the 3D simulation, we see the larger plume growing relative to the smaller, leaving a dominant driving plume. This is similar to the behavior in 2D. A persistent wind that emerges $\sim$300 ms after bounce is present in both the large and small explosion plumes, and finally in the dominant plume alone. We see simultaneous explosion and accretion $-$ the smaller plume in Fig. \,\ref{fig:2} growing relatively in size. Even up to the end of our simulation, some material partially circumnavigates the explosion plumes, plunges inward in a sheet that seems to pinch off the larger from the smaller plume, and is accreted onto the PNS. This accretion pinching in the early explosion phase between the two differently-sized exploding plumes resembles a wasp's waist and may be a common feature of some CCSN explosion morphologies. The smaller plume is more prominent in 3D than in 2D, for which at late times the opposing explosion plume's volume ratio is significantly smaller than in 3D.

An inner structure with two counter-ejected large lobes such as we see in this simulation, with one demonstrably larger than the other, crudely resembles the iron ejecta pattern inferred from XMM X-ray observations of the supernova remnant (SNR) Cas A (\citealt{willingale}). This is suggestive, but the remnant structure in any SNR depends upon its entire propagation history through the star's matter field and any apparent morphological association between early and late ejecta patterns could be happenstance. This remains to be determined. However, the rough similarity between our preliminary debris field morphology and the inferred inner mass density and composition patterns from X-ray data is indeed intriguing.


The 3D simulation has lower maximum entropies at late
times by $\sim$4.5 units (Boltzmann constant (k$_B$) per baryon) than its 2D counterpart. However, the entropy averaged over the shocked region (defined where the specific entropy is greater than 4 k$_B$ per baryon) is comparable for both simulations. This is because the 3D simulation maintains a more `isotropic' explosion in that even the subdominant plume subsists, producing comparable mean entropies over the shocked region despite the higher entropies along the dominant axial plume in 2D.




\subsection{Shock Wave Evolution}\label{sec:shock}

\begin{figure*}
\includegraphics[width=0.49\textwidth]{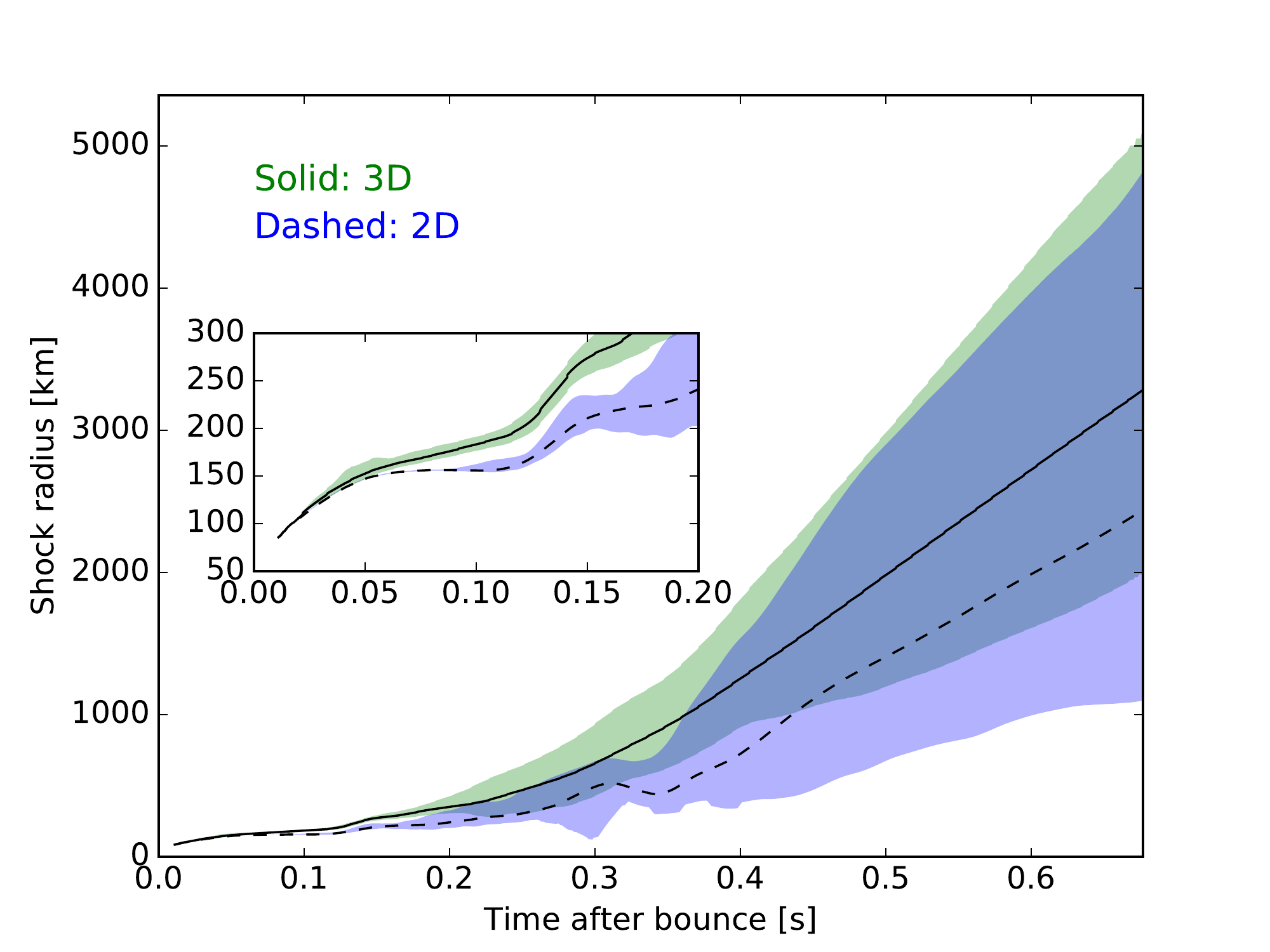}
\hfill
\includegraphics[width=0.49\textwidth]{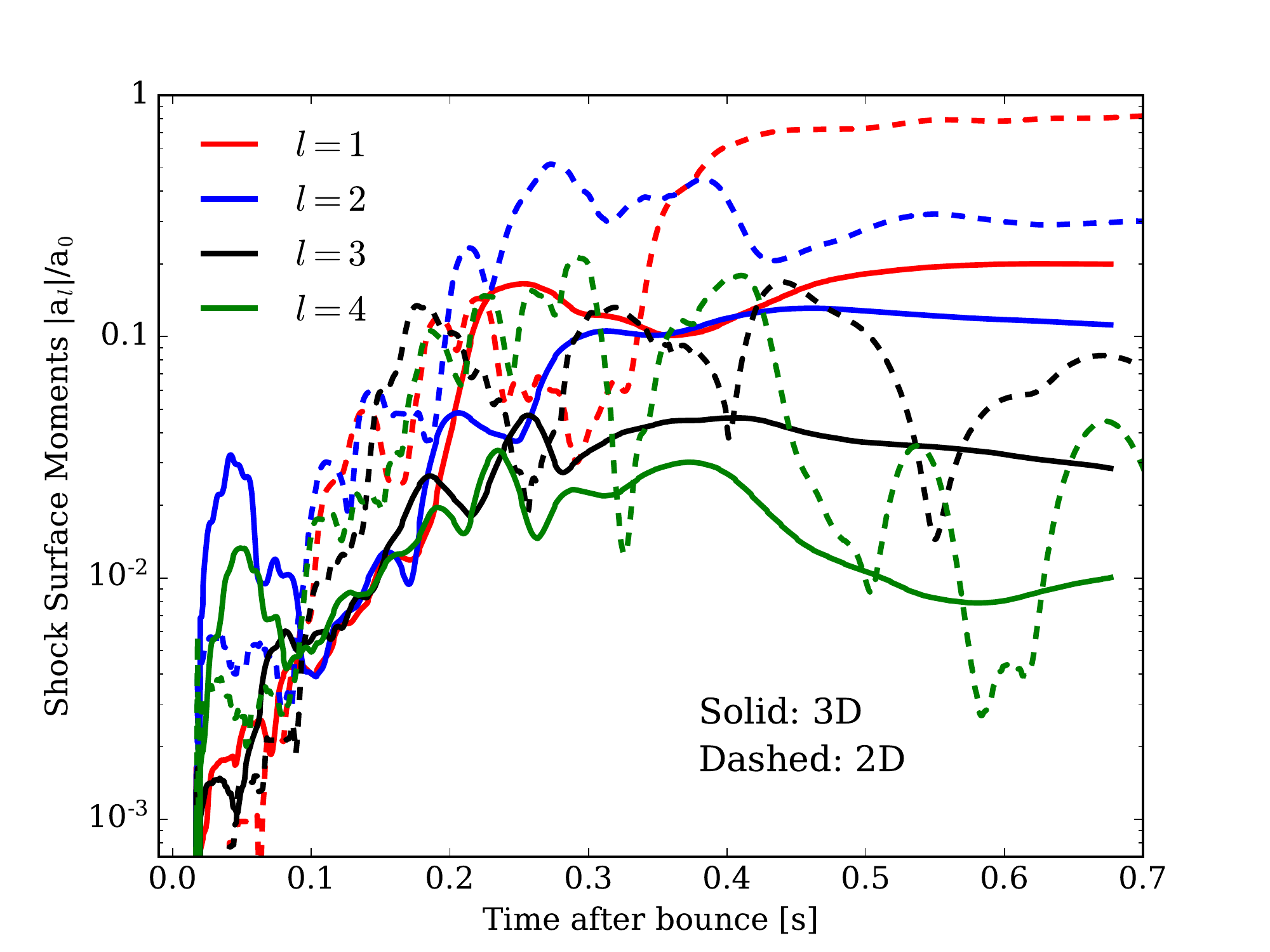}
\caption{\textbf{Left}: The shock radius (km) vs. time after bounce (in seconds) for the 2D
(dashed, blue swath) and 3D (solid, green swath). The colored-in regions indicate the range
of the shock location, from minimum to maximum. The 3D simulation explodes slightly earlier.
At the end of our simulation, the shock achieves $\sim$5000 km. The shock of the 3D model
barely stalls in radius, while the shock for its 2D counterpart stalls for $\sim$50 ms.
We show in the inset a zoomed-in plot of the average shock radii at early times. The
mean shock radii for the 2D and 3D simulations have diverged by $\sim$50 ms after bounce.
\textbf{Right}: The first four spherical harmonic moments of the shock radius as a function of time (in seconds)
after bounce, normalized to the mean shock radius (the $\ell = 0$ component). We take
the norm over all orders $m$ and compare 3D (solid) to 2D (dashed). Up to $\sim$70 ms after bounce,
the $\ell=2,4$ moments dominate, the former due to the initial quadrupolar velocity perturbations
imposed. From $\sim$100 to $\sim$200 ms, all reduced moments are comparable in magnitude.
At late times, the large scale, lower-$\ell$ moments increase in significance. Up to $\ell=11$
(not shown), we find monotonically decreasing relative moment magnitudes with increasing $\ell$
(and decreasing angular scale). We see a transition from small structures at early times
to large structures at later times. Up to explosion, the 3D simulation evinces much larger deviations from spherical symmetry. At late times, however, the 2D simulation shows much larger
asymmetries than the 3D simulation, indicated by the larger magnitude of the reduced moments.}
\label{fig:3}
\end{figure*}

We find, perhaps surprisingly (\citealt{2012ApJ...755..138H,2013ApJ...770...66H,2013ApJ...765..110D}) that our 3D model explodes
roughly 50 ms earlier than the corresponding 2D model. In the top panel of Fig.\,\ref{fig:3},
we plot the shock radius versus time after bounce for the 2D (dashed, blue swath)
and 3D (solid, green swath). The colored-in areas indicate the radial spread of the shock
location, from minimum to maximum. At the end of our 3D simulation, the mean shock radius has reached beyond
$\sim$5000 km. The 2D model remains roughly spherical in expansion for the first $\sim$120 ms,
whereas the 3D model deviates from spherical symmetry earlier. We show in the inset
a zoomed-in plot of the average shock radii at early times. The shock radii for the 2D
and 3D simulations diverge around $\sim$50 ms after bounce. The shock of the 3D model barely
stalls, while the shock for its 2D counterpart stalls for $\sim$50 ms.

In the bottom panel of Fig.\,\ref{fig:3}, we plot the first four spherical harmonics of the shock radius
as a function of time after bounce. We take the norm over all orders $m$ and compare 3D
(solid) with 2D (dashed). We use the approach outlined in \cite{burrows2012} to
decompose the shock surface R$_s(\theta,\phi)$ into spherical harmonic components with coefficients:
\begin{equation}
a_{lm} = \frac{(-1)^{|m|}}{\sqrt{4\pi(2l+1)}} \oint R_s(\theta,\phi) Y_l^m(\theta,\phi) d\Omega\, ,
\end{equation}
normalized such that $a_{00} = a_{0} =\langle R_s\rangle$ (the average
shock radius) and $a_{11}$, $a_{1{-1}}$, and $a_{10}$ correspond to the average
Cartesian coordinates of the shock surface $\langle x_s\rangle$, $\langle y_s\rangle$,
and $\langle z_s\rangle$, respectively.
The orthonormal harmonic basis functions are given by
\begin{equation}
Y_l^m(\theta,\phi) = \begin{cases}
        \sqrt{2} N_l^m P_l^m(\cos\theta) \cos m\phi&            m>0\, ,\\
        N_l^0 P_l^0(\cos\theta) &                               m=0\, ,\\
        \sqrt{2} N_l^{|m|} P_l^{|m|}(\cos\theta) \sin |m|\phi&  m<0\, ,
\end{cases}
\end{equation}
where
\begin{equation}
N_l^m = \sqrt{\frac{2l+1}{4\pi}\frac{(l-m)!}{(l+m)!}}\, ,
\end{equation}
$P_l^m(\cos\theta)$ are the associated Legendre polynomials,
and $\theta$ and $\phi$ are the spherical coordinate angles. We plot the norm,

\begin{equation}
P_{\ell} = \frac{\sqrt{\sum_{m=-\ell}^{\ell} a_{\ell m}^2}}{a_{00}}\,.
\end{equation}

Up to $\sim$70 ms after bounce, the $\ell=2,4$ moments dominate, the former due to the
initial quadrupolar velocity perturbations imposed. From $\sim$100 to $\sim$200 ms, all moments
are comparable in magnitude. Note that the dip in the quadrupole moment at $\sim$300 ms
corresponds to the dip in mean shock radius seen in the left panel. Shortly afterwards, the shock
surface of the 2D simulation rapidly expands, catching up with that of the 3D simulation. At
late times, the large-scale, lower $\ell$ moments dominate. Up to $\ell=11$ (not shown), we
find that the moment magnitudes decrease monotonically with increasing $\ell$ (and decreasing
angular scale). We witness a transition from small structures at early times, coalescing into
large-scale structures at later times.  As the explosion commences, the 3D simulation evinces
larger deviations from spherical symmetry than the 2D simulation, as indicated by the larger
magnitudes of the respective moments. However, at late times the 2D simulation begins to manifest
larger asymmetries than its 3D counterpart, indicated by the larger magnitude of the lower-order
moments. Both simulations have similar asymptotic shock velocities and maximum shock radii
(at a given post-bounce time), though the 2D simulation
minimum and average shock radii are roughly $\sim$1000 km smaller.

In Fig.\,\ref{fig:5}, we
track the dipole orientation of the shock with time. Early on, the shock dipole vector
changes sporadically (but does not simply jump up and down as in 2D), but at later times it settles to
an axis seemingly chosen arbitrarily. The randomly chosen axis is a defining feature of 3D
non-rotating simulations (see, e.g. Fig.\,3 in \citealt{burrows2012}).
We also see pronounced azimuthal structures in the 3D simulation (as opposed to rings in 2D).
Along with the $\ell = 0$ explosion mode, the $\ell=1, m=-1$ dipolar mode dominates
at late times, and we see such a structure in the 3D explosion maps (Fig.\,\ref{fig:1}).

\subsection{Energetics}\label{sec:energy}

Before explosion, the energy deposited in the gain region, that thick shellular volume
interior to the shock wave where  neutrino heating rate exceeds the cooling rate, is most relevant
for driving turbulence and establishing the potential for explosion.  The larger the
energy deposition rate, the closer a given progenitor model  is (with its mass accretion 
rate) to explosion (\citealt{1993ApJ...416L..75B}).  However, the total energy deposited 
in advance of explosion is not related to the explosion energy 
(\citealt{bhf1995}). The matter heated in the gain region is subsequently advected into the PNS,
where it first reradiates a fraction of the acquired energy and then merges with the radiating PNS.
It is only after the explosion commences that the deposited energy might be retained to contribute
to the asymptotic explosion energy.  However, even though explosive expansion leads to diminished
cooling as the matter temperatures decrease, there continues to be some neutrino cooling.
More importantly, the exploding matter expands against gravity, so that much of the ongoing neutrino
energy deposited is used to lift the matter out of the deep potential well.  This explains why the neutrino
heating rates even during explosion are larger than the accumulation rate of the supernova blast 
energy in the first seconds of the explosion phase.  Recombination of the nucleons into nuclei
will provide a boost ($\sim$9 MeV per baryon) to the asymptotic kinetic energy of the supernova ejecta,
but the associated recombining mass is generally not large (here $\sim$0.08 M$_{\odot}$). Moreover, 
the associated total energy is comparable to the gravitational binding term.  As a result, it 
appears that the supernova will take many seconds to achieve its final energy.  Therefore, even 
though our 3D simulation of this 16-M$_{\odot}$ progenitor's core has been conducted farther 
post-bounce than any other simulation with state-of-the-art numerics and microphysics, 
we have captured only the early stages of an explosion that will need to be followed 
numerically for a few more seconds to witness the asymptoting of the explosion energy (\citealt{muller2016, muller2017}).

\begin{figure*}
\includegraphics[width=0.49\textwidth]{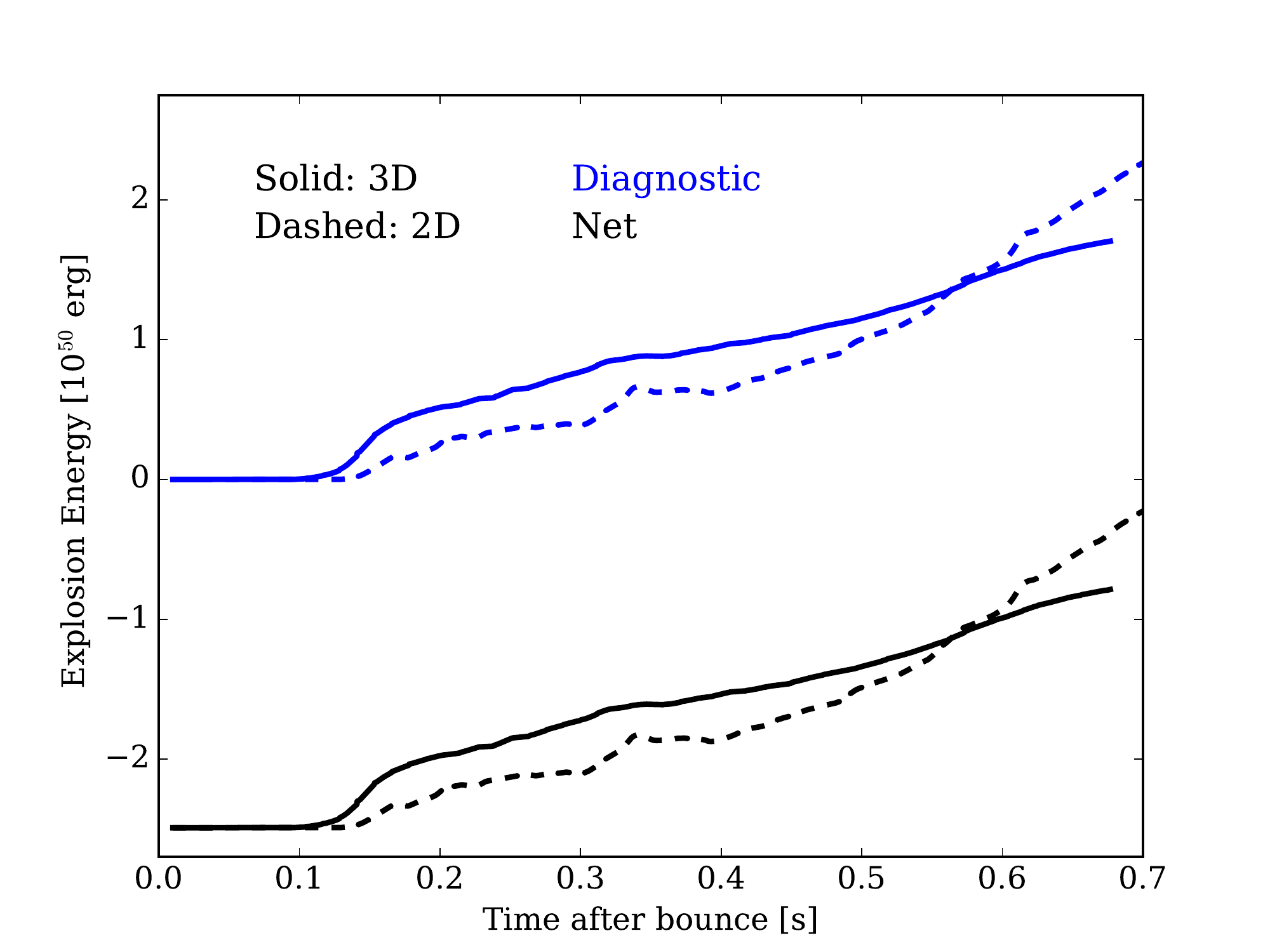}
\hfill
\includegraphics[width=0.49\textwidth]{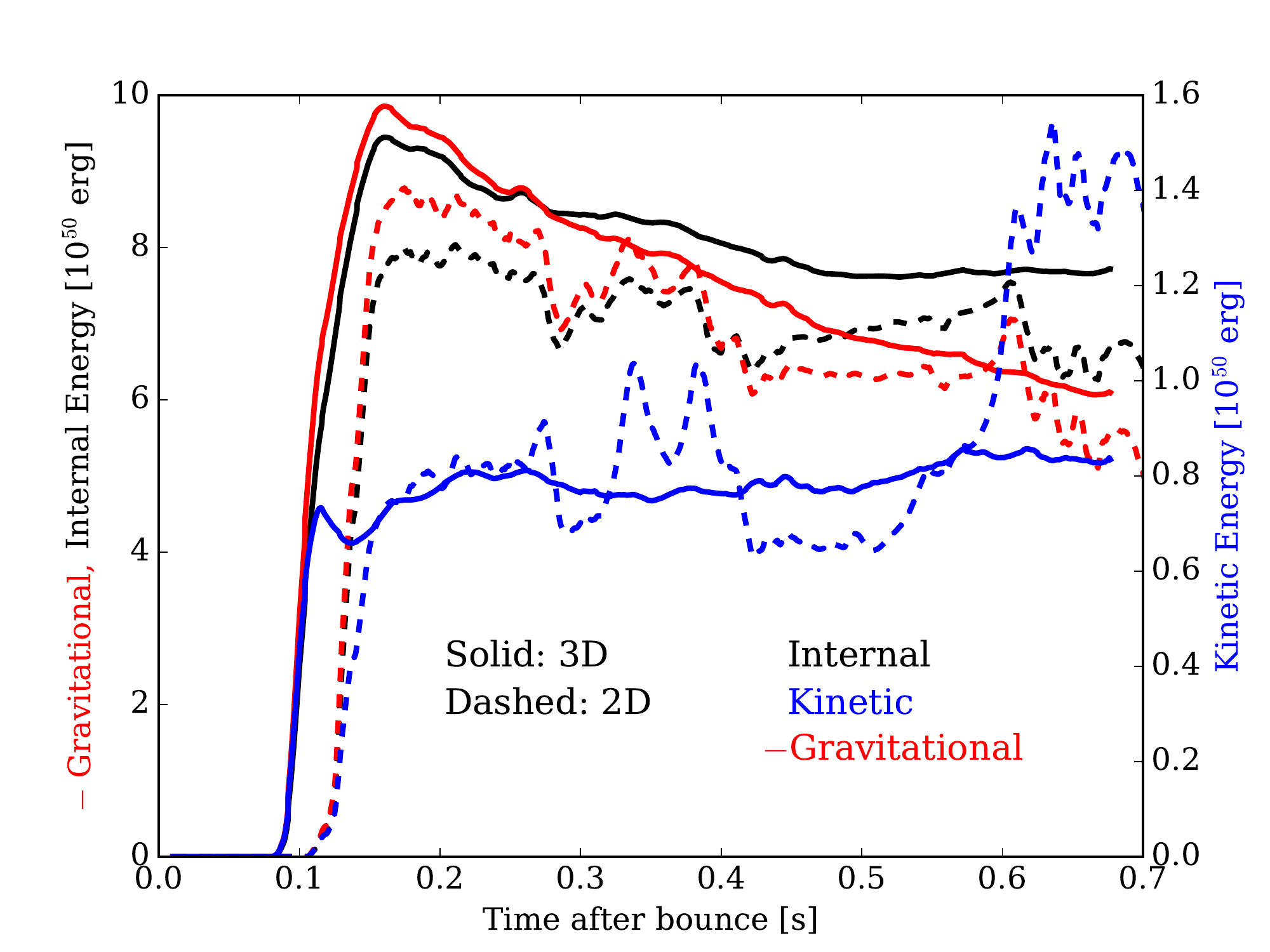}
\caption{\textbf{Left}: Diagnostic (blue) and net (black) explosion energies (in 10$^{50}$ erg) for 
the 16-M$_{\odot}$ progenitor as a function of time after bounce (in seconds). \textbf{Right}: 
Internal (blue, left y-axis) and kinetic (green, right y-axis) energies (in 10$^{50}$ erg) as 
a function of time after bounce (in seconds). Solid indicates the 3D model and dashed the corresponding 2D model 
for both figures. The diagnostic energy (left, green) does not account for the gravitational 
overburden of $\sim$2.5$\times 10^{50}$ erg exterior to our simulation grid (outer boundary 10,000 km). 
The total explosion energy (blue) is not yet positive for the 3D simulation (at 677 ms after bounce), 
though the 3D simulation explodes slightly earlier. The 3D simulation maintains a higher 
internal energy, by $\sim$15\%, through the end of the simulation, and a higher explosion energy, 
but similar kinetic energies, until $\sim$550 ms after bounce. The subsequent rise in explosion energy 
for the 2D model corresponds with the steep rise in its kinetic energy, also seen 
in \protect\cite{vartanyan2018a} for the same 16-M$_{\odot}$ progenitor, but with a different initial setup. 
Such a sharp rise in kinetic energy is not seen in our 3D simulation.}
\label{fig:5}
\end{figure*}

The total energy we plot in Fig. \ref{fig:5} is comprised of the kinetic energy, the thermal energy, 
the recombination energy, and the gravitational energy of the ejecta.  The so-called ``diagnostic"
energy ignores the binding energy (thermal plus gravitational) of the progenitor exterior 
to the computational domain.  Here, the total energy quoted includes this penalty, different 
for every progenitor and outer computational boundary radius; including this term is required 
to assess the true supernova explosion energy.

We calculate diagnostic energies for our 16-M$_{\odot}$ progenitor in 3D and 2D, summing the kinetic, thermal, gravitational, and nuclear binding energies interior to our 10,000-km simulation grid where the matter parcel's Bernoulli term is positive. We correct for the gravitational binding energy of $2.5\times10^{50}$ erg exterior to our grid, and plot both the diagnostic (blue) and net (black) explosion energies in the left panel of Fig.\,\ref{fig:5}. In the right panel, we plot the thermal (blue, left y-axis), gravitational (red, left y-axis) and kinetic (green, right y-axis) energies (in 10$^{50}$ erg) as a function of time after bounce (in seconds). Solid indicates the 3D model and dashed the 2D analog for both figures.

The 3D model explodes slightly earlier and initially has a higher explosion energy than its 2D model counterpart (Fig.\,\ref{fig:5}). At the end of the simulation, 677 ms after bounce, the 3D model has a diagnostic explosion energy of 1.7$\times$10$^{50}$ erg. Accounting for the gravitational overburden, the total explosion energy (blue) is not yet positive for the 3D simulation (-0.8$\times10^{50}$ erg).  Before $\sim$550 ms after bounce, the 3D simulation maintains similar kinetic energies and a higher internal energy by $\sim$15\% than its 2D analog. Thenabouts, the 2D model explosion energy overtakes that of the 3D model, with the rise in explosion energy corresponding to a steep rise in its kinetic energy at a growth rate of $\sim$5$\times10^{50}$ erg s$^{-1}$. Such a rise, also at $\sim$550 ms after bounce, is  seen in \protect\cite{vartanyan2018a} for the same 16-M$_{\odot}$ progenitor, but with a different initial setup. It is not seen in our 3D simulation.  We conjecture that the stronger dipole and quadrupole moments of the 2D simulation (Fig.\,\ref{fig:3}, right) relative to those of the corresponding 3D model at late times contribute to this divergence in kinetic energy. At the end of our simulation, the explosion energy for the 3D model is climbing at a rate of approximately 2.5$\times$10$^{50}$ erg s$^{-1}$, half that of the 2D case. Similar energy growth rates are found for the 3D simulations in the literature (see, e.g. \citealt{muller2017}) and necessitate continuing 3D simulations for several seconds. 

\begin{figure}
\includegraphics[width=0.41\textwidth]{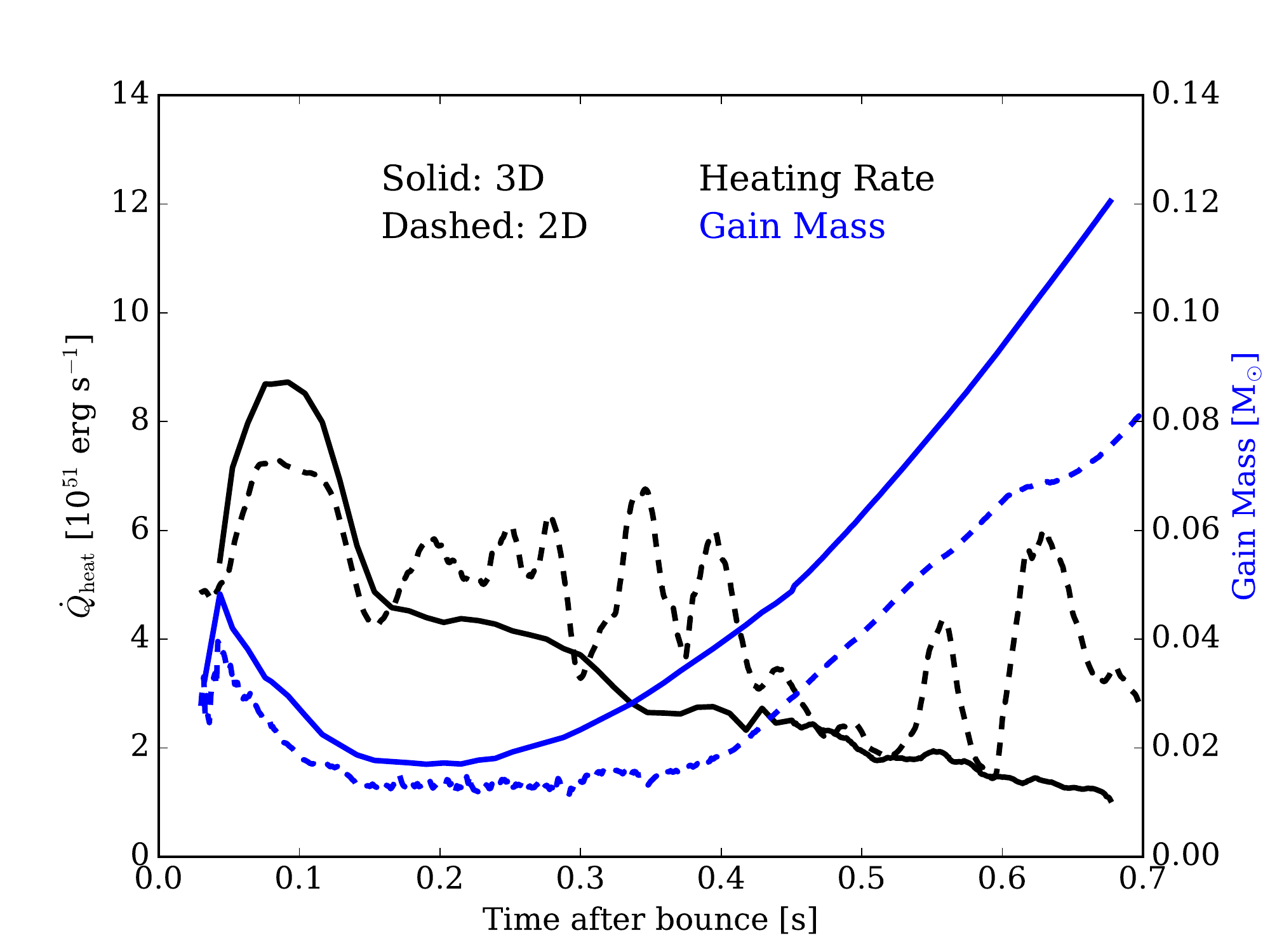}
\includegraphics[width=0.41\textwidth]{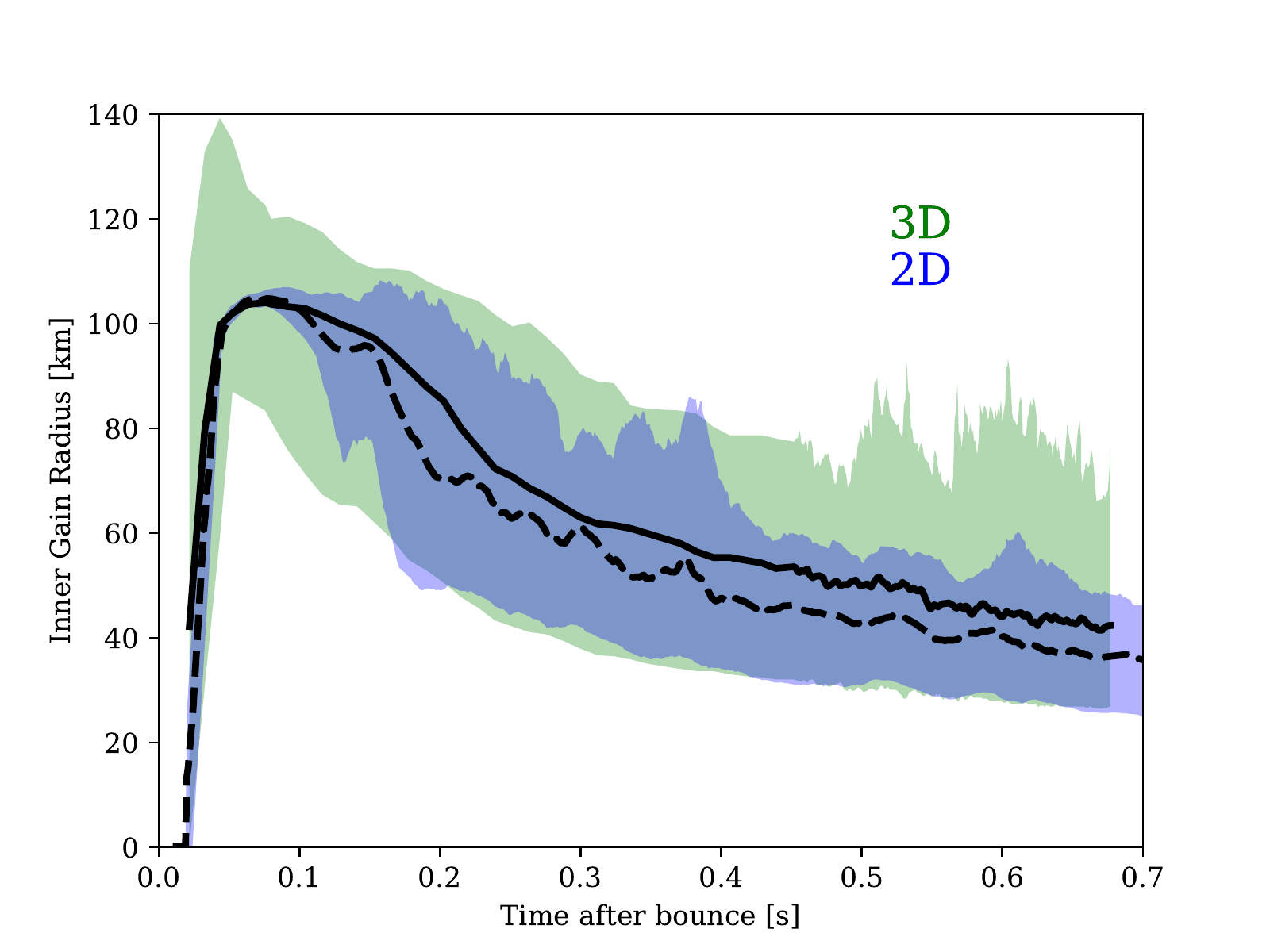}
\includegraphics[width=0.41\textwidth]{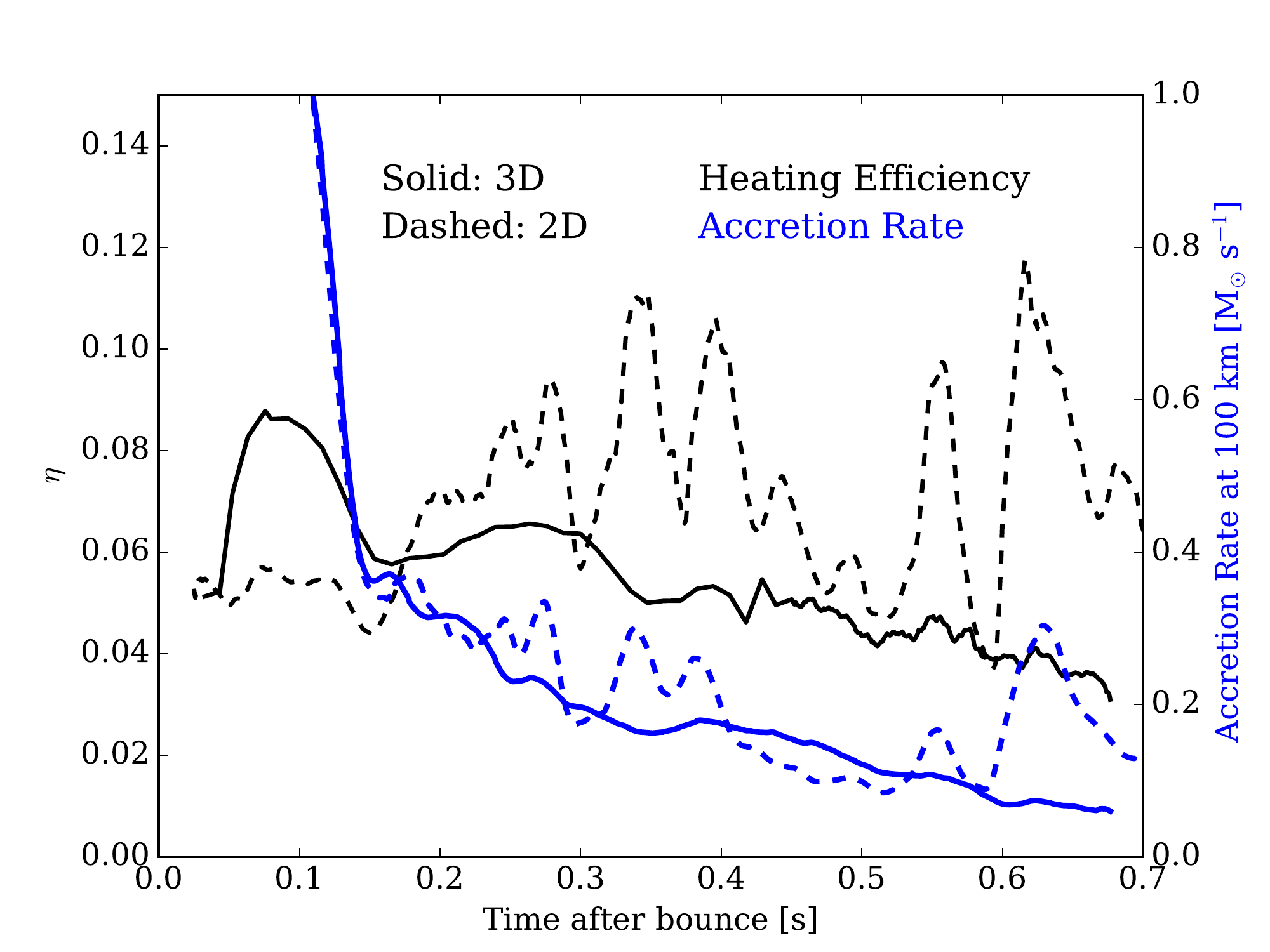}
\caption{\textbf{Top}: We illustrate the heating rates (blue, 10$^{51}$ erg s$^{-1}$), and the gain 
mass (black, in 10$^{-3}$ M$_{\odot}$) as a function 
of time after bounce (in seconds) for the 3D (solid) and 2D (dashed) simulations of the 16-M$_{\odot}$ 
progenitor. Prior to explosion ($\sim$100 ms), the heating rate for the 3D simulation is $\sim$30\% higher than for the 2D simulation. The gain mass is also slightly higher for 
the 3D model, exceeding 0.12 M$_{\odot}$ at the end of our simulation.  
\textbf{Middle}: Inner boundary of the gain region (in km) as a function of time after bounce (in seconds). Black lines depict the mean positions of the inner gain region (solid for 3D, dashed for 2D). The 3D 
simulation (green, solid) maintains a much larger variation of the inner boundary of the gain region throughout the evolution.
\textbf{Bottom}: Heating efficiency $\eta$ (black), defined as the gain-region heating rate divided by the 
sum of the $\nu_e$ and $\bar{\nu}_e$ luminosities entering the gain region, and the accretion rate at 150 km (blue, in M$_{\odot}$ s$^{-1}$). Through the first $\sim$150 ms, 
the 3D simulation (green) has a heating efficiency $\sim$40\% higher than the 2D (blue) simulation. 
However, after $\sim$200 ms, the 2D simulation overtakes the 3D simulation, and showcases a high degree 
of variability over $\sim$50-ms time scales. Note the correlation between jumps in accretion rate and jumps 
in heating rates (and efficiencies) in the 2D simulation. }
\label{fig:6}
\end{figure}

In Fig.\,\ref{fig:6}, top panel, we illustrate the heating rates and the gain mass as a function of time after bounce for the 3D (solid) and 2D (dashed) simulations of the 16-M$_{\odot}$ progenitor. Just prior to explosion, at $\sim$100 ms, the heating rate for the 3D simulation is $\sim$30\% (2 Bethe s$^{-1}$) higher than for the corresponding 2D model. The gain mass is also slightly higher for the 3D model, exceeding 0.12 M$_{\odot}$ at the end of our simulation).  After $\sim$150 ms post-bounce, we see more variability in the heating rate for the 2D simulation than for the 3D simulation. Through almost $\sim$700 ms after bounce, the growth rate of the explosion energy is less than 20\% of the heating rate, the difference due to the work done against gravity by the ejecta. It is not until late times that the growth rate of the explosion energy is expected to be close to the heating rate. 
In the middle panel, we show the spread of the inner boundary of the gain region a function of time after bounce, defined here where the net heating (heating minus cooling) is greater than 10\% of the heating alone. The 3D simulation (green, solid) maintains a much larger variation in radial boundary throughout the evolution, extending almost twice as far at late times as the 2D model. In the bottom panel, we show the heating efficiency $\eta$, defined as the heating rate divided by the luminosity entering the gain region,

\begin{align}
      \eta = &\frac{\dot{Q}_{heat}}{L_{\nu_e}+L_{\bar{\nu}_e}}\,.
\end{align}

Through the first $\sim$150 ms, the 3D simulation has a heating efficiency of $\sim$0.09, 40\% higher than the corresponding 2D model. However, after 200 ms, the efficiency of the 2D simulation overtakes that of the 3D simulation, and showcases a high degree of variability with a time scale of $\sim$50 ms.

\begin{figure}
\includegraphics[width=0.5\textwidth]{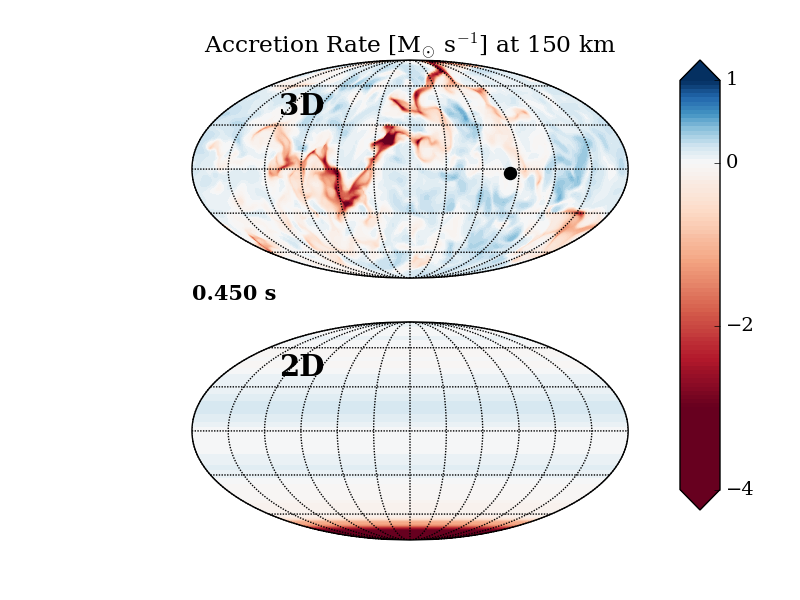}
\caption{Mollweide projections of the accretion rate for the 3D and the 2D simulations at 450 ms after bounce. The spatial variation of accretion rate in the 3D simulation is in sharp contrast with the accretion rate in the 2D simulation, where we see only a dominant dipole component in the southern hemisphere.}
\label{fig:lum_moll}
\end{figure}

In Fig.\,\ref{fig:lum_moll}, we show Mollweide projections of the accretion rate for the 3D and 2D models. The spatial variations for the 3D simulation for the accretion rate contrast sharply with that for the 2D simulation, in which we see a dominant dipole component only in the southern hemisphere. 


\subsection{Luminosity and Mean Energies}\label{sec:lum}

In Fig.\,\ref{fig:lum}, we plot the luminosity (top) and mean energies (bottom) at a radius of 500 km as a function of time after bounce. Note that the luminosities and average energies for the 2D and 3D models are remarkably similar and show significant difference only beyond $\sim$600 ms after bounce. We note, however, key differences in the electron-neutrino luminosities through the first $\sim$150 ms, with the 2D simulation boasting a luminosity $\sim$7\% larger than that for the 3D simulation. Furthermore, the `heavy'-neutrino luminosity is $\sim$3\% smaller for the 2D simulation than for the 3D simulation over the same time period. We explore this more in Sec.\,\ref{sec:pns}. Here, we remark that the interplay between the greater electron-neutrino luminosity and the smaller `heavy' neutrino luminosity in the critical first one-hundred ms of our 2D simulation (compared to our 3D model) impede earlier explosion revival in the 2D case. The former strips the gain region of energy deposition by neutrinos (since the electron-type neutrinos have a much higher absorption opacity than the `heavy'-type neutrinos). Furthermore, the greater `heavy'-neutrino luminosity in the 3D simulation may act in the same direction as the axial-vector many-body correction to produce a harder electron-neutrino spectrum and facilitate explosion (\citealt{burrows2018}). The culmination of these effects is visible in Fig.\,\ref{fig:6}, top panel, where a small difference in the respective luminosities translates into a significantly smaller heating rate in the 2D simulation compared to the analogous 3D simulation.



\subsection{Ejecta Composition}\label{sec:ejecta}

Our calculations follow the evolution in space and time of the electron fraction, Y$_e$.
This quantity is an essential determinant of subsequent nucleosynthesis.  While we do not in this 
paper derive the detailed elemental composition of our ejecta, the distributions of 
the entropies and Y$_e$s in the inner explosion debris provide qualitative information 
on the likely character of the emergent element burden.  In our previous 2D simulations 
(\citealt{vartanyan2018a}), histograms of the ejecta Y$_e$ were derived.  What 
we found was that much of the ejecta have Y$_e$s above 0.5, implying that the 
ejected matter has been processed by differential $\nu_e$ and $\bar{\nu}_e$ absorption 
that has made some of it proton-rich. This is what we witness in this 3D simulation,
though whether this is a generic outcome remains to be determined.  Proton-rich ejecta 
could be a site of the p- and $\nu$p-processes (\citealt{pruet2006,frohlich2006,wanajo2011})
and might be the context for the production of some of the first peak of the r-process 
(\citealt{hoffman1996,pruet2006,frohlich2006,wanajo2011,frebel2018,bliss2018}).

\begin{figure}
\includegraphics[width=0.5\textwidth]{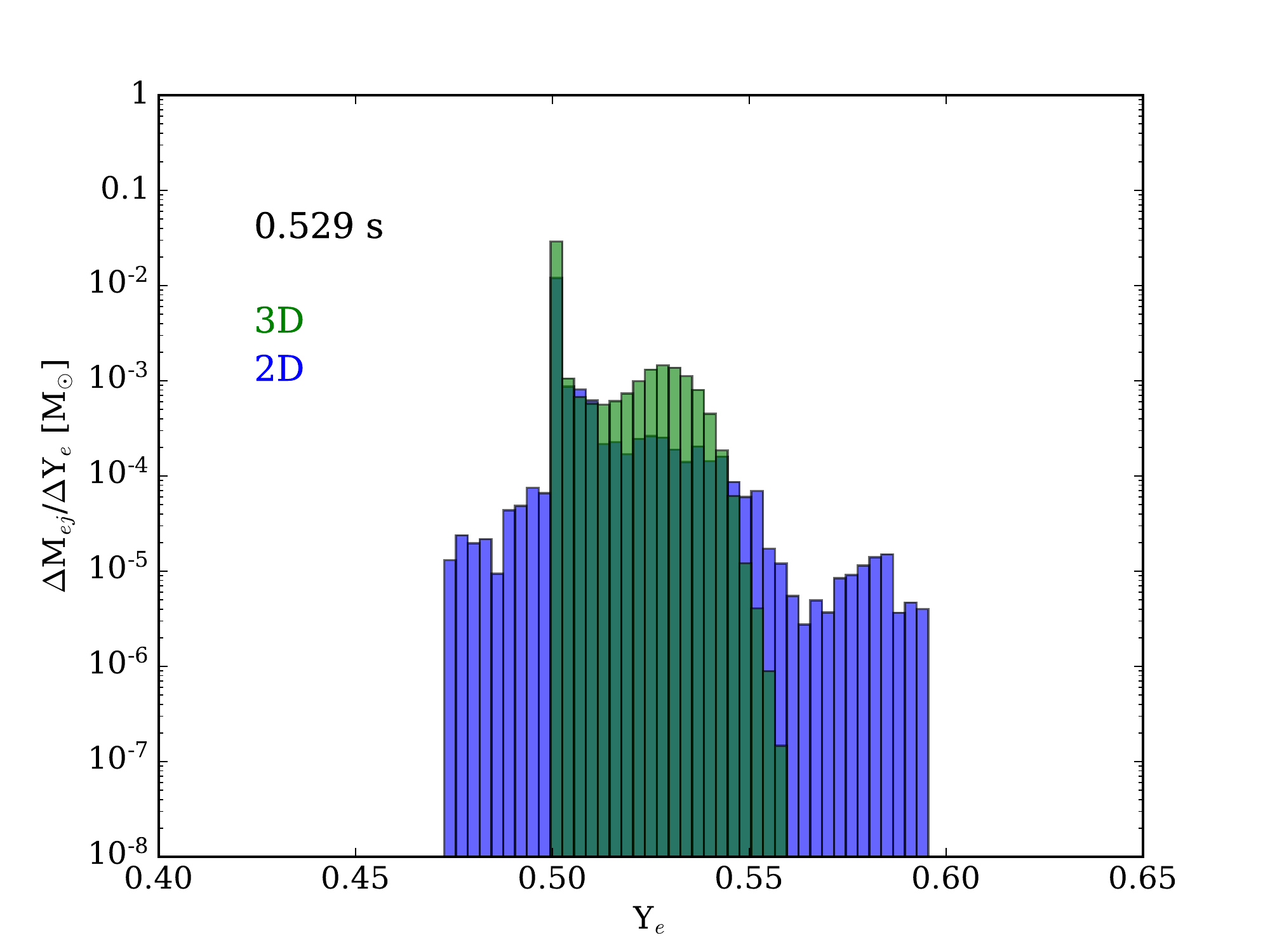}
\caption{Histogram of ejecta mass distribution by Y$_e$ at 0.529 seconds) after bounce. The green bars 
indicate the results of the 3D simulation, and the blue those for the 2D simulation. We find the 
interesting result that the ejecta mass distribution in 2D has a tail extending out to both higher 
($>0.55$) and lower ($<0.5$) Y$_e$ than the 3D simulation at a given time.}
\label{fig:hist_Ye}
\end{figure}

In Fig.\,\ref{fig:hist_Ye}, we provide a histogram of the ejecta mass distribution in Y$_e$ at 0.529 seconds after bounce. The green bars indicate the results of the 3D simulation, and the blue bars that of the 2D simulation. Though both models peak at Y$_e$ = 0.5, we find the interesting result that the ejecta distribution in the 2D model has a wider tail extending out to both higher ($>0.55$) and lower ($<0.5$) Y$_e$ than the 3D simulation at any given time. For much of the evolution, the ejecta of the 3D simulation spans Y$_e$ $\sim$0.5$-$0.55, whereas the ejecta in the 2D simulation encompasses Y$_e$$\sim $0.45$-$0.6. Only at late times does the 3D simulation have significant low-Y$_e$ ejecta at large radii (see the violet tail in Fig.\,\ref{fig:ye_vis}). \footnote{We provide here the Y$_e$ distribution in ejecta (defined as gravitationally unbound mass) beyond 1,000 km. We also looked at the Y$_e$ mass distribution of ejecta beyond 100 km. Our conclusion that our 3D simulation has a narrower Y$_e$ span than the 2D model, remains unchanged.}

In an earlier paper on 2D models (\citealt{vartanyan2018a}), we found that only the 16-M$_{\odot}$ progenitor had an ejecta-Y$_e$ distribution that extended to lower Y$_e$, among the four progenitors considered. We claimed that an anisotropic explosion, with much of the outflow directed toward one hemisphere, would leave the opposite hemisphere with relatively untouched neutron-rich material. We see a similar result here. The 3D simulation, on the other hand, produces a more omnidirectional explosion $-$ leaving little matter untouched. The achievement of higher Y$_e$ in 2D can similarly be understood $-$ the concentration of explosion in one direction in the 2D simulation allows ample neutrino processing of the ejecta to higher Y$_e$. 

\begin{figure*}
\includegraphics[width=0.45\textwidth]{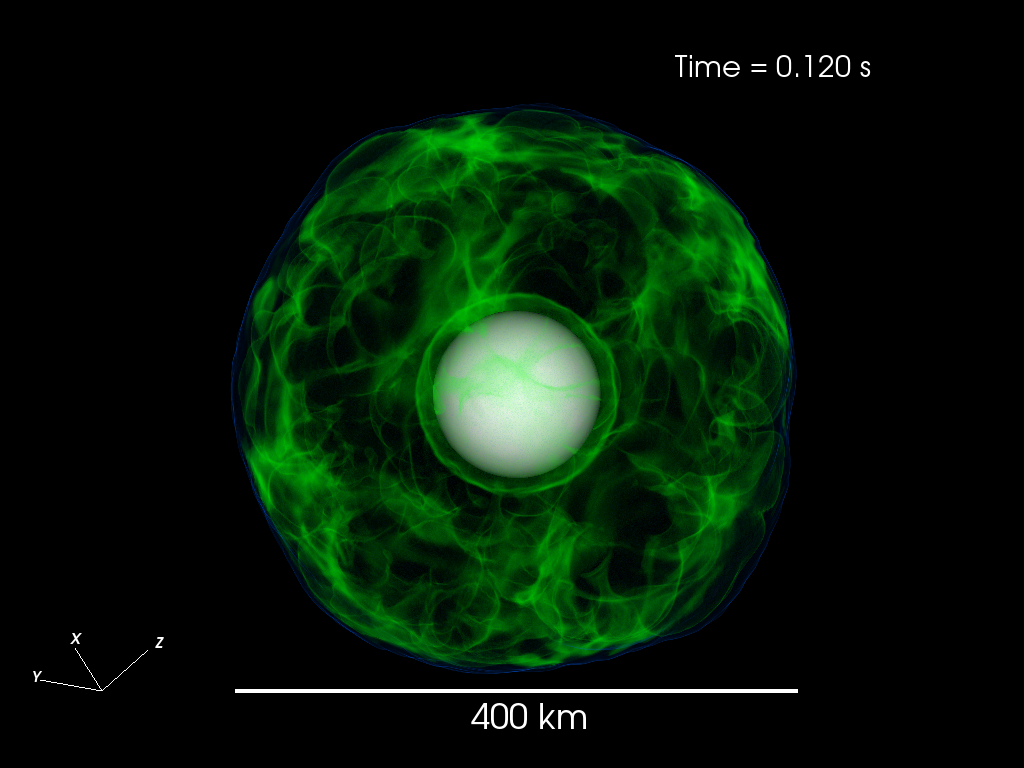}
\includegraphics[width=0.45\textwidth]{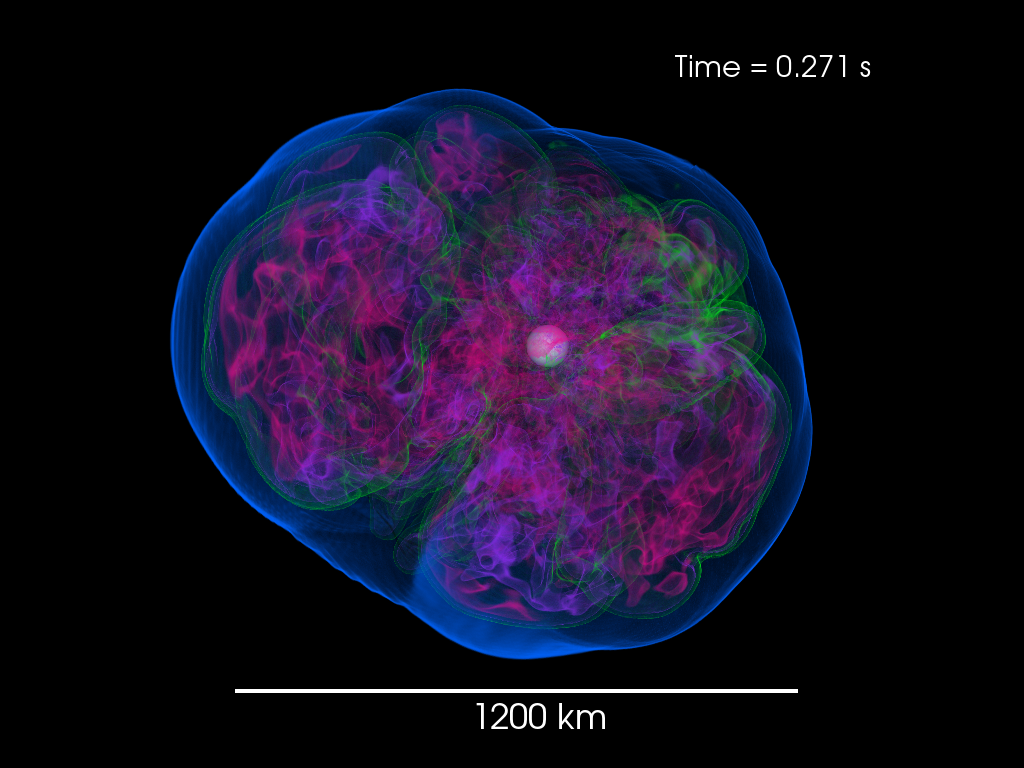}
\includegraphics[width=0.45\textwidth]{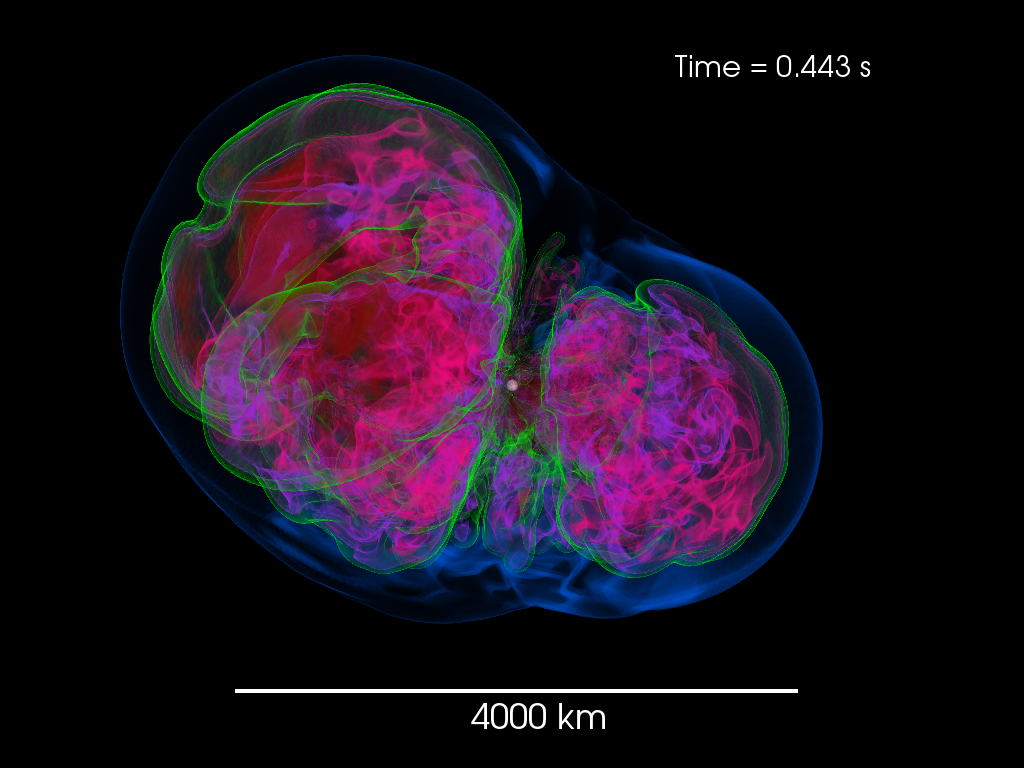}
\includegraphics[width=0.45\textwidth]{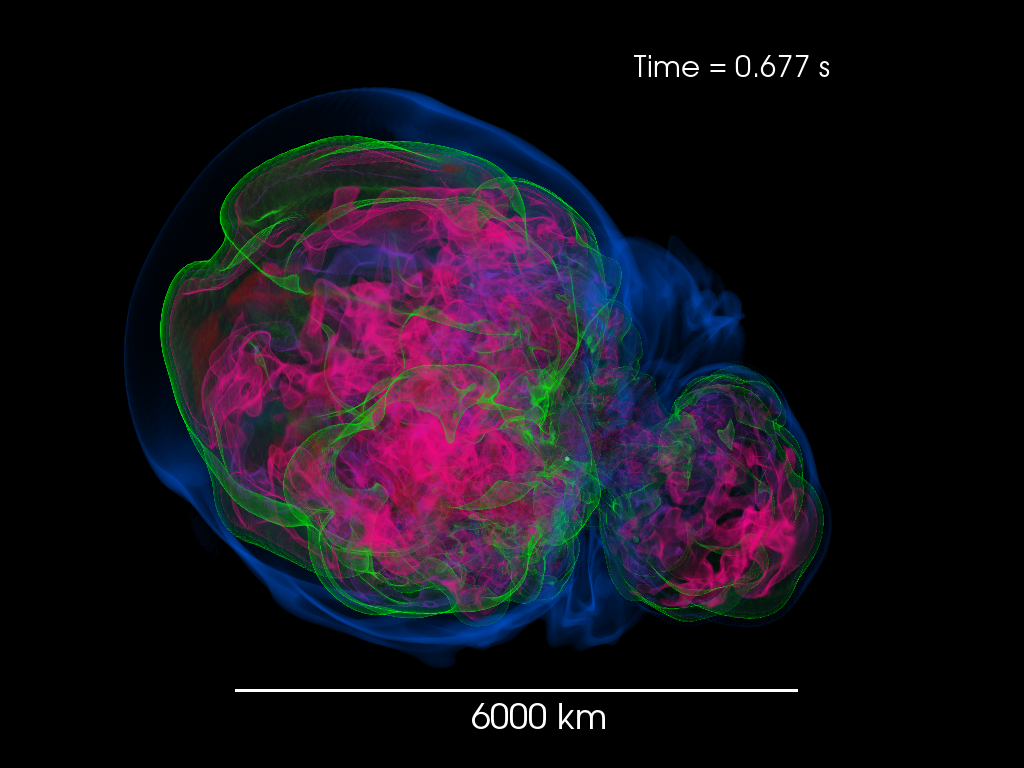}
\caption{Time sequence of the entropy of the 16-M$_{\odot}$ progenitor. Note the different spatial scales. The 
inner white sphere is a 10$^{11}$ g cm$^{-3}$ isosurface that roughly delineates the PNS, and the blue veil traces 
an entropy contour of 4-k$_b$/baryon, a proxy for the shock radius. Note the bifurcated cerebral 
structure of the explosion plumes, with one dominant hemisphere (on the left in this projection). 
Several ``fingers" are also visible along the axis, though these are accreted shortly after. Note the 
high-entropy regions (dark red) both along the outer cusps of the plumes and in the interior as matter is funneled 
onto the PNS.}
\label{fig:1}
\end{figure*}

\begin{figure}
\includegraphics[width=0.49\textwidth]{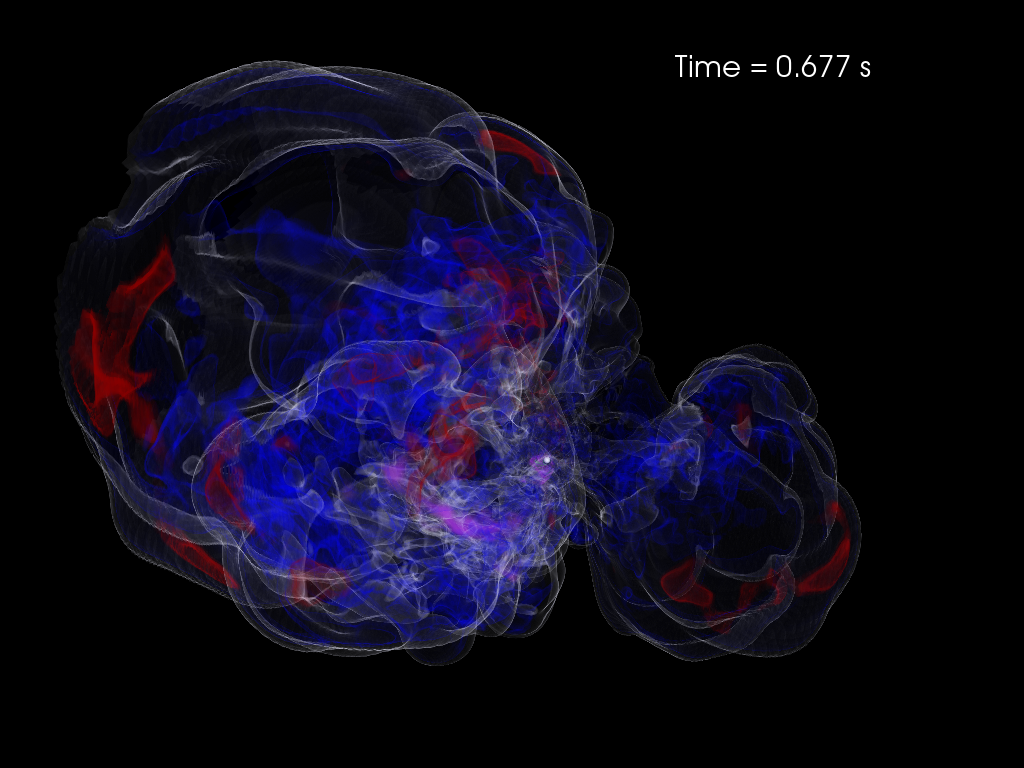}
\caption{Y$_e$ distribution at $\sim$677 ms after bounce. The white ``veil" encompasses the 
expanding plumes, just interior to the shock radius, at a Y$_e$ of 0.5. The blue plumes 
indicate a Y$_e$ spanning the interval 0.5 - 0.52, and the red caps a Y$_e$ greater than 0.52. The 
latter is concentrated along the exterior cusps of the plumes, and interior where accretion is 
funneled onto the PNS. Note the resemblance of the high-Y$_e$ distribution to the entropy 
distribution in Fig.\,\ref{fig:1}. The violet tail shows the low-Y$_e$ ($<$ 0.5) ejecta seen 
in Fig.\,\ref{fig:hist_Ye}. This trailing `tail' is also visible in the density evolution of the progenitor.}
\label{fig:ye_vis}
\end{figure}

We illustrate the 3D distribution of Y$_e$ in the ejecta in Fig.\,\ref{fig:ye_vis} at $\sim$667 ms after bounce. The white ``veil" illustrates a Y$_e$ of 0.5, just interior to the location of the shock radius. The high-Y$_e$ plumes correspond to the high-entropy plumes of Fig.\,\ref{fig:1}, with the blue plumes indicating Y$_e$'s that span 0.5 $-$ 0.52, and the red blobs Y$_e$ greater than 0.52. The latter is concentrated along the exterior cusps of the plumes, and in the interior where accretion is funneled onto the PNS. Note the resemblance of the high-Y$_e$ distribution in Fig.\,\ref{fig:ye_vis} to the entropy distribution in Fig.\,\ref{fig:1}.





\subsection{Inner PNS Convection}\label{sec:pns}

The original delayed explosion mechanism of \cite{wilson1985} was facilitated by the enhancement of 
the driving neutrino luminosities by what he termed ``neutron-finger" convection.
This was a doubly-diffusive instability, akin to salt-finger convection in the oceans,
that was suggested to result in an otherwise stably-stratified PNS.  A Ledoux-stable 
balance of Y$_e$ and entropy gradients was thought to be undermined by the 
more rapid diffusion of energy vis \`a vis lepton number.  Wilson captured this effect in 
1D spherical models of explosion with a mixing-length-like diffusive flux, and the $\nu_e$ 
and $\bar{\nu}_e$ luminosities were thereby augmented by $\sim$25\%.  Without this
effect, Wilson's models did not explode.  However, \cite{bruenn_dineva}
showed that the core was not unstable to such ``neutron-finger" convection, and this was
confirmed by \cite{2006ApJ...645..534D} using 2D simulations.  However, after bounce, there
is a region in the PNS between $\sim$10 and $\sim$30 kilometers that is in fact unstable
to classical convection, driven mostly by negative Y$_e$ gradients.  This PNS convection
is a feature in all modern multi-dimensional simulations of CCSN.  In their study, \cite{2006ApJ...645..534D}
noticed that this overturning convection increased the emergent luminosities, but by the end
of their simulation $\sim$200$-$300 ms after bounce this increase was not large.
In addition, inner PNS convection and the outer neutrino-driven convection interior to the stalled 
shock did not merge into one large convective zone. Given this, \cite{2006ApJ...645..534D}
concluded that PNS convection was not centrally important to the neutrino mechanism of
CCSNe.

On the contrary, in their study of the lowest-mass progenitor stars, \cite{2017ApJ...850...43R} found
that the contribution of a PNS convection boost to the emergent neutrino luminosities
grew with time after bounce, and could reach significant fractions.  This was particularly 
true for $\bar{\nu}_e$ and $\nu_{\mu}$ neutrinos, for which the respective 
neutrinospheres are deepest.  Here, we explore the corresponding effects
and numbers for our 3D simulation of the 16-M$_{\odot}$ progenitor of \cite{wh07}, 
and compare them to the 2D case.  

\begin{figure}
\includegraphics[width=0.41\textwidth]{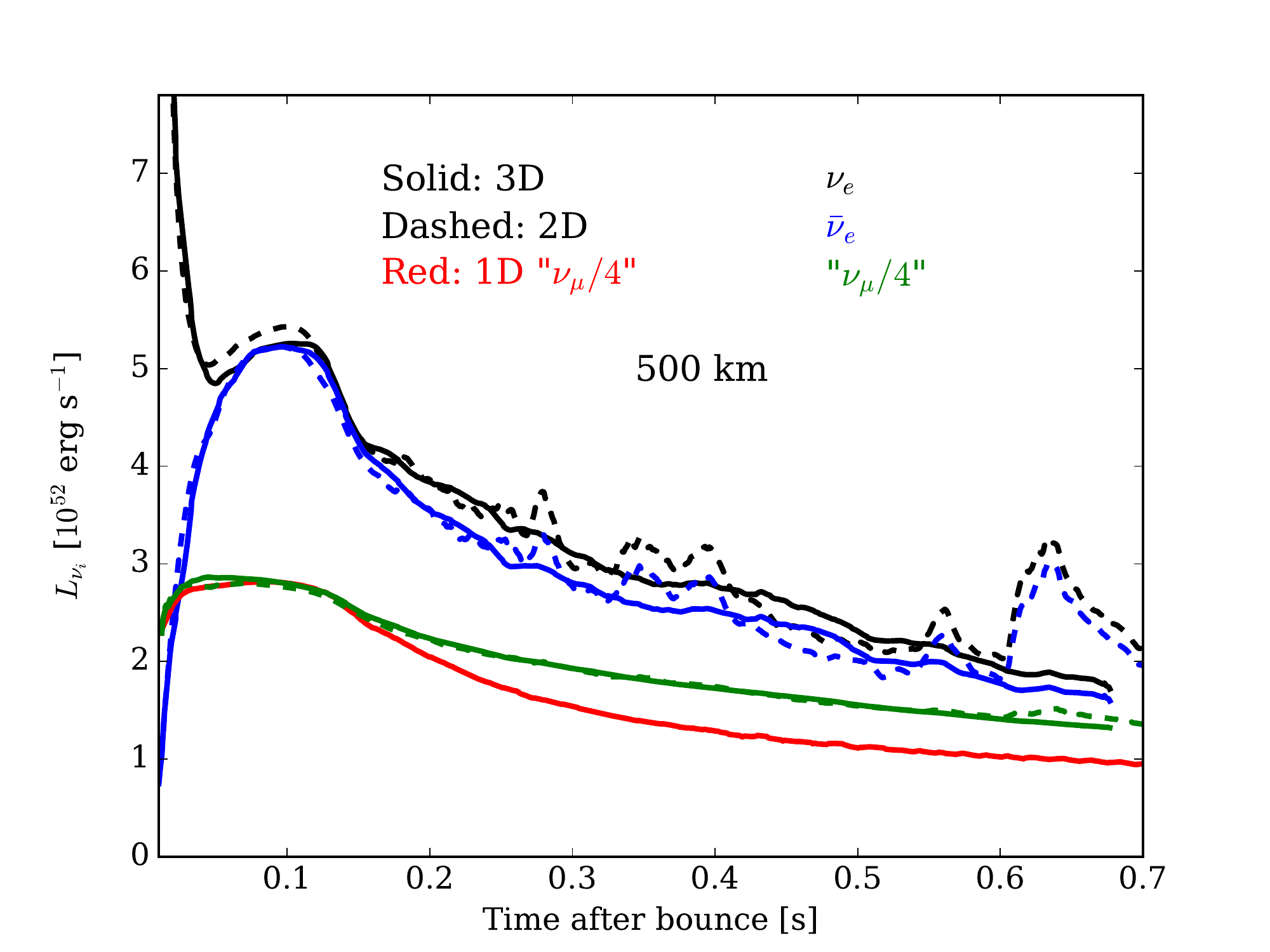}
\hfill
\includegraphics[width=0.41\textwidth]{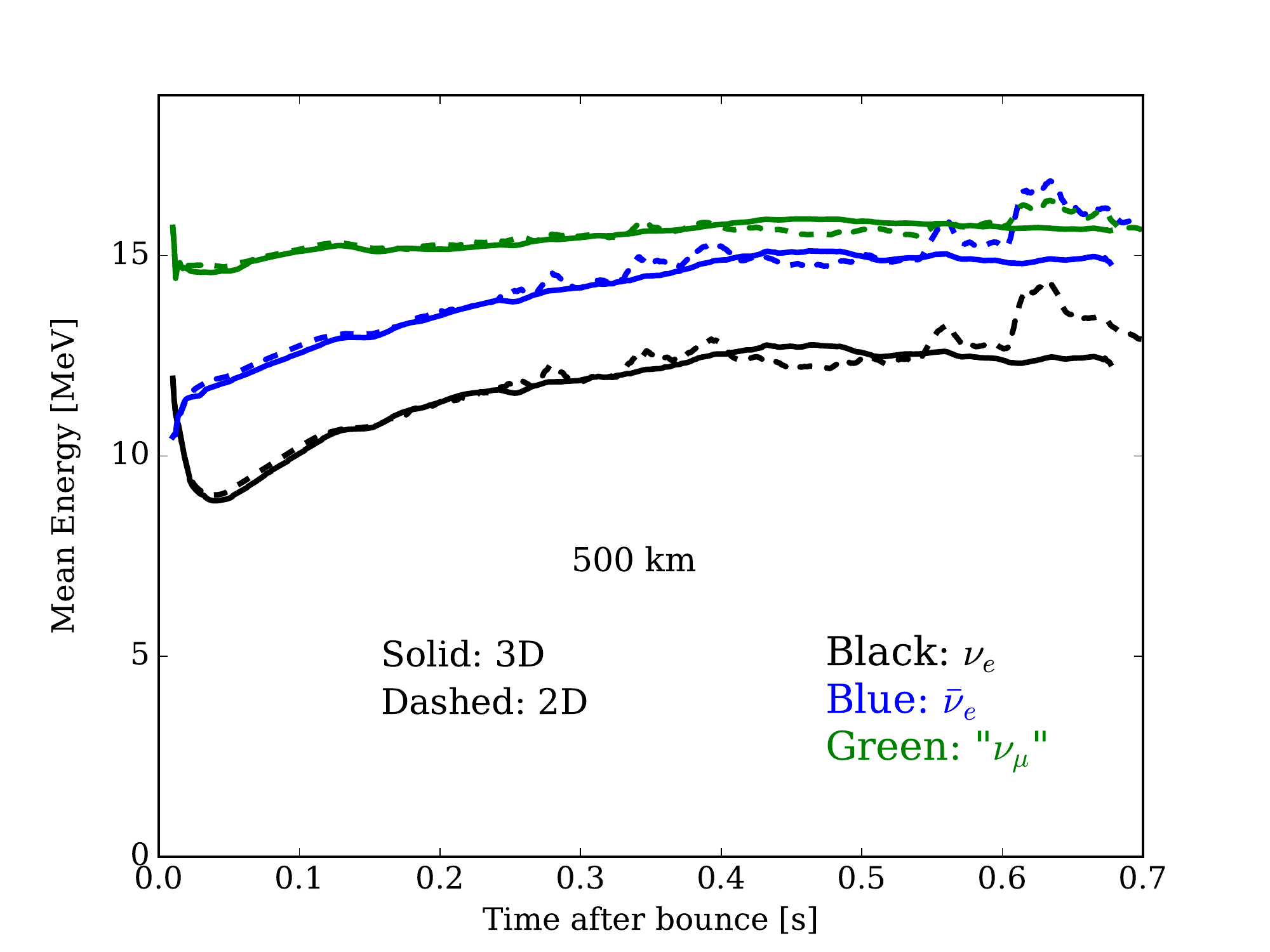}
\caption{Neutrino luminosity (\textbf{top}, 10$^{51}$ erg s$^{-1}$), and average neutrino energy (\textbf{bottom}, MeV) 
as a function of time after bounce (in seconds) at 500 km. Note that the luminosities and average 
energies for 2D and 3D are remarkably similar and show a significant difference only after 600 ms after bounce.}
\label{fig:lum}
\end{figure}

\begin{figure}
\includegraphics[width=0.5\textwidth]{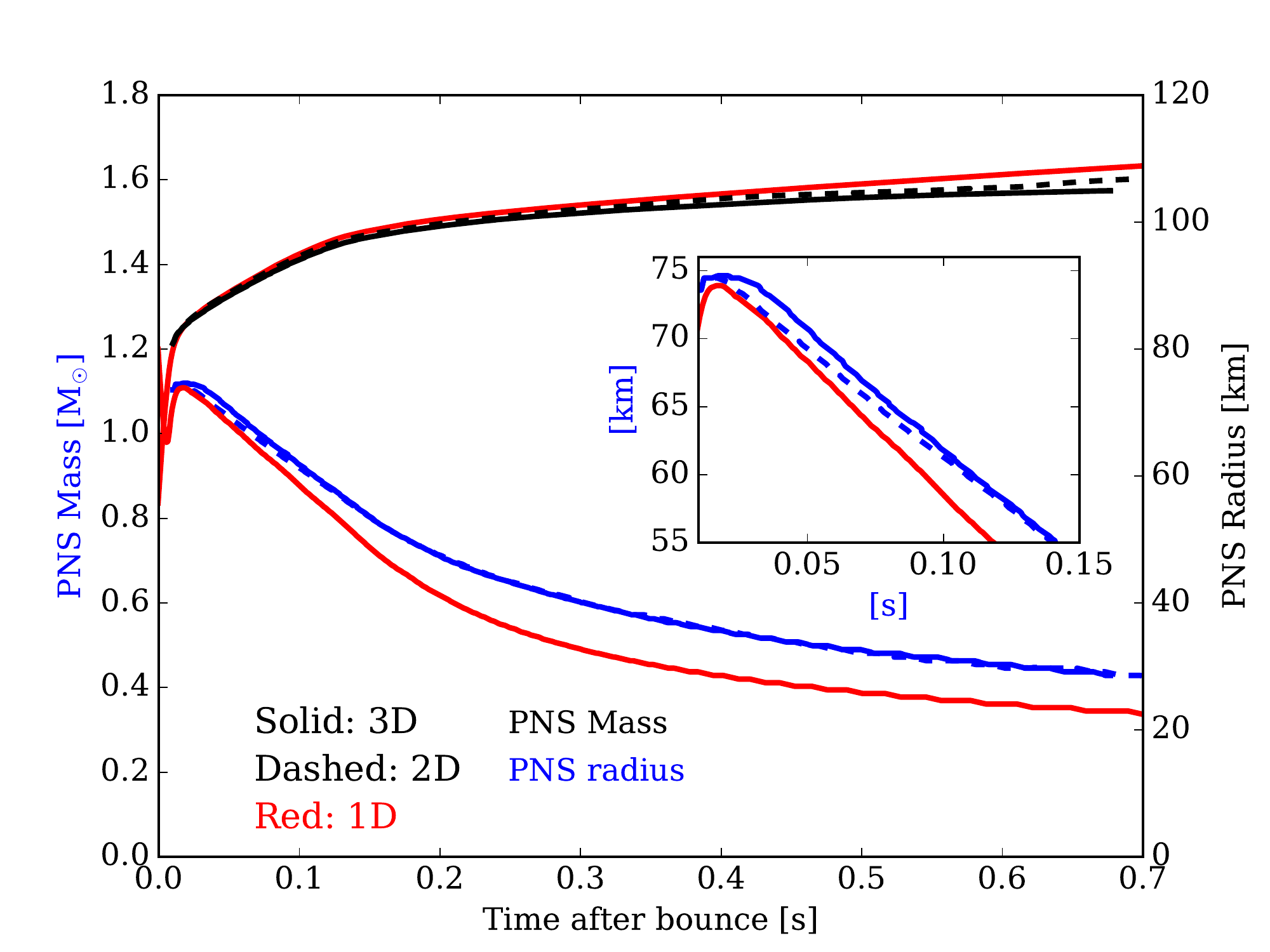}
\caption{The PNS mass (in M$_{\odot}$, blue) and radius (in km, black) as a function of time after 
bounce (in seconds) for the 3D (solid), 2D (dashed), and 1D (red) simulations of the 16-M$_{\odot}$ progenitor. 
At late times, the PNS radii for the 2D and 3D simulations are virtually identical, but significantly smaller 
in the 1D case. The larger PNS mass in 1D than 2D, and in 2D than 3D, is due to the longer accretion history 
than in 3D, where we see early explosion. In the inset, we show the PNS radius zoomed in for the first 150 ms 
after bounce. Until $\sim$140 ms after bounce, the PNS radius in the 2D simulation is as much as $\sim$3\% smaller than for the 3D simulation.}
\label{fig:pns}
\end{figure}

\begin{figure*}
\includegraphics[width=0.49\textwidth]{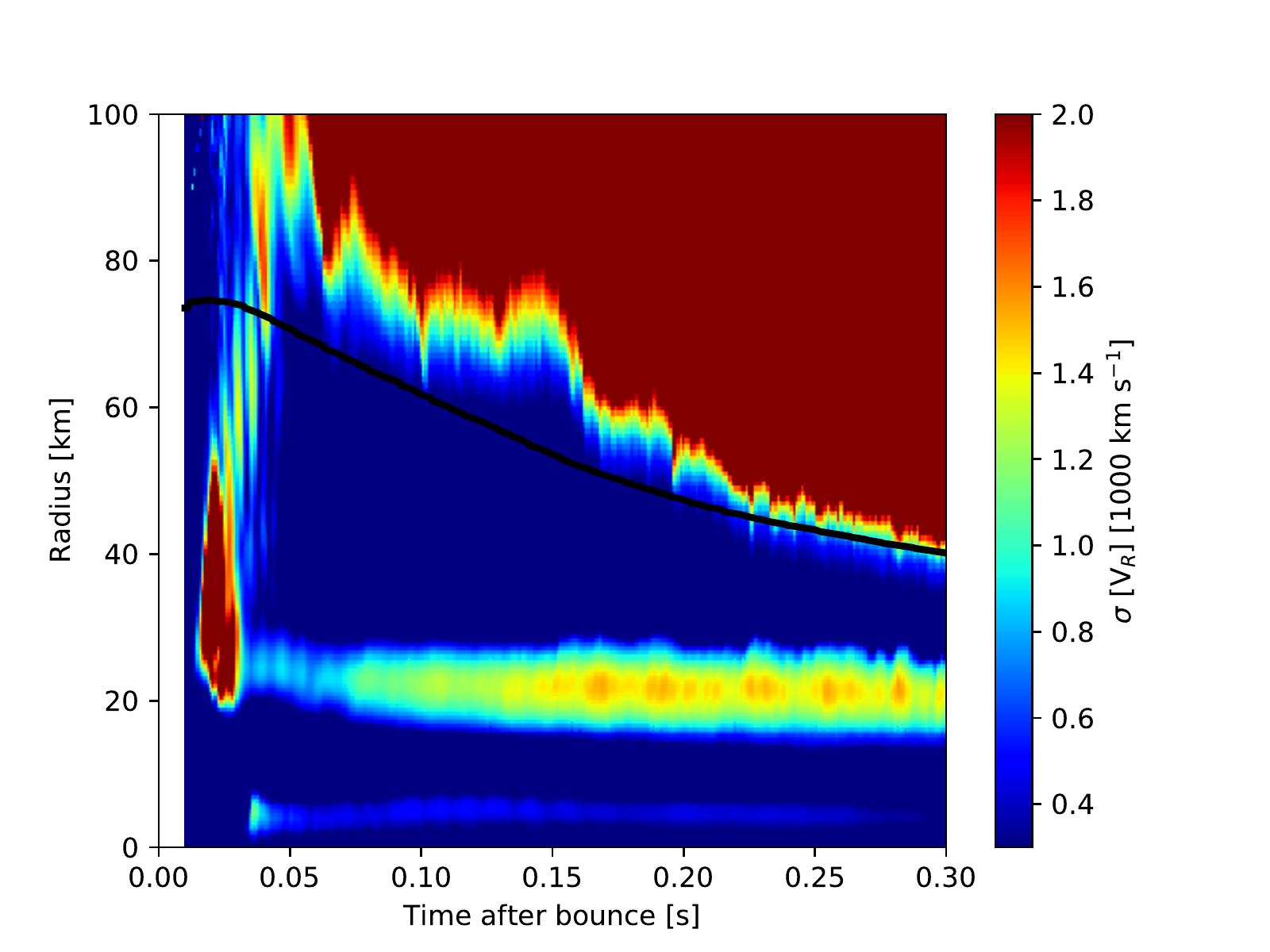}
\hfill
\includegraphics[width=0.49\textwidth]{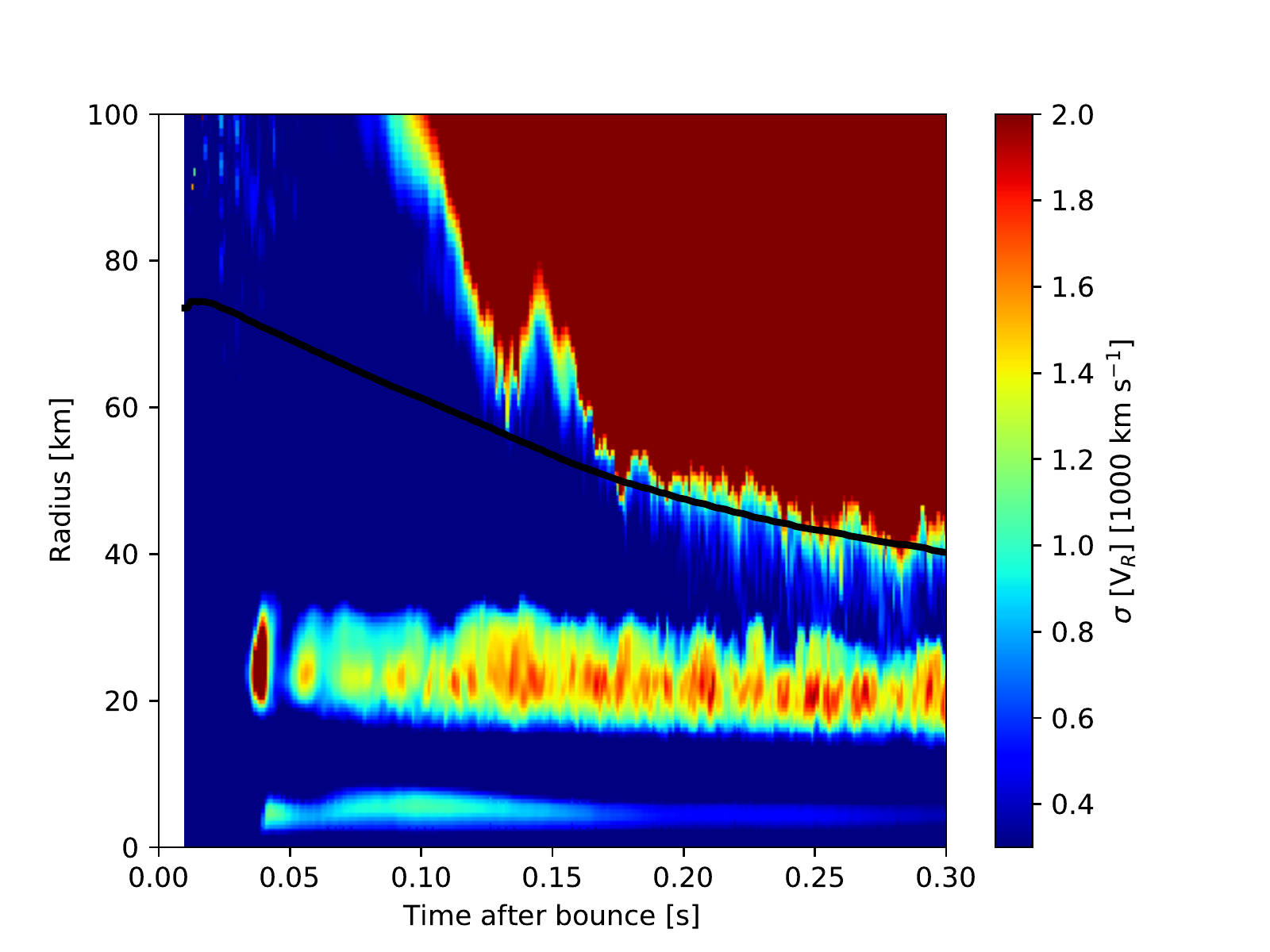}
\caption{A space-time diagram of the standard deviation $\left(\sqrt{\langle (v_r - \langle v_r\rangle)^2\rangle}\right)$ over 
angle of the radial velocity within the inner 100 km through 300 ms after bounce for the 3D (left) 
and 2D (right) models.  Note that it is significantly smaller in 3D than in 2D (see also Fig.\,\ref{fig:ye_slide}). 
Both the outer and inner (PNS) convective regions are visible here, and the interior convective zone is a band in 
velocity similar to that seen in \protect\cite{2006ApJ...645..534D}. The black lines illustrate the mean PNS radius, which in 3D, and not 2D, is sampled by the outer neutrino-driven convection through the first 120 ms after-bounce. By $\sim$300 ms after bounce, the exterior convective 
zone has receded to $\sim$50 km. In the 2D simulation, the interior convective zone is a few km wider and has 
higher convective velocities by several hundred km s$^{-1}$ than its 3D counterpart. Furthermore, we see more 
variation in the radial location of the convective zones in the 2D simulation, with the outer convective zone making 
excursions almost to the inner convective zone by $\sim$300 ms after bounce.  See the text for further discussion.
}
\label{fig:convec_cmap}
\end{figure*}

We plot in Fig.\,\ref{fig:pns} the PNS mass (in M$_{\odot}$, blue) and mean radius (black) as a function of time after bounce for the 3D (solid), 2D (dashed), and 1D (red) simulations of the 16-M$_{\odot}$ progenitor. The PNS surface here is defined where the density is 10$^{11}$ g cm$^{-3}$. The baryonic PNS mass in our 3D simulation at $\sim$677 ms after bounce is $\sim$ 1.57 M$_{\odot}$ (1.6 M$_{\odot}$ in the 2D model, 1.63 M$_{\odot}$ in the 1D model), corresponding to a gravitational mass of 1.42 M$_{\odot}$ (1.44 M$_{\odot}$ in the 2D model, 1.47 M$_{\odot}$ in the 1D model). The PNS mass reflects the disruption of net accretion onto the PNS. Interestingly, we find that the difference between the PNS mass for the 1D and 2D models is roughly comparable to the difference in the same quantity between the 2D and the 3D models at late times, despite the absence of explosion in the 1D model and the correspondingly lengthier accretion history. Furthermore, at late times, the mean PNS radii in the 2D and 3D simulations are virtually identical ($\sim$29 km) but are significantly smaller in the 1D case ($\sim$23 km). A similar dependence of the PNS radii on simulation dimension was found in \cite{2017ApJ...850...43R} and \cite{vartanyan2018a}. Here, we have the opportunity to compare such quantities to that of a 3D simulation. In the inset, we show the PNS radius zoomed in for the first 150 ms after bounce. Until $\sim$140 ms after bounce, the PNS radius in the 2D simulation is as much as $\sim$3\% smaller than in the 3D  model, lying between the PNS radii in the 3D simulation and in the 1D simulation. Simultaneously, as shown in Fig.\,\ref{fig:lum}, the ``heavy''-neutrino luminosity is slightly smaller in the 2D simulation than in the 3D simulation. At later times, both the ``heavy''-neutrino luminosity and the PNS radius in the 1D simulation are significantly lower than in the multidimensional simulations (see also \citealt{radice2017a,vartanyan2018a}). On time scales greater than $\sim$200 ms, PNS convection boosts the ``heavy''-neutrino luminosities in the 2D and 3D simulations. Furthermore, the shrinking PNS radius comes into close contact with the inner convective region after 200 ms (see Fig.\,\ref{fig:convec_cmap}), explaining the larger PNS radii in multi-dimensional simulations.  However, electron-type neutrino luminosities are higher in 1D than in multidimensional simulations simply because that model does not explode, and accretion power remains significant. 

In Fig.\,\ref{fig:convec_cmap}, we provide a space-time diagram of the standard deviation over angle of the radial velocity within the inner 100 km through 300 ms after bounce for the 3D (left) and 2D (right) models. Both convective regions are visible here as the bright regions $-$ the interior convective band is similar to that seen in \protect\cite{2006ApJ...645..534D}, and the exterior, neutrino-driven convection recedes to $\sim$50 km by $\sim$300 ms. The interior convective zone in the 2D simulation is a few kilometers wider and has higher convective velocities than its 3D counterpart. Furthermore, we see more variation in the radial location of the convective zones in the 2D simulation. However, in the 3D simulation, the exterior, neutrino-driven convective region is located deeper in at early times, reaching $\sim$80 km by $\sim$50 ms after bounce in the 3D simulation (by comparison, the exterior convective region in the 2D case reaches 80 km more than 100 ms after bounce). Through the first $\sim$150 ms, this exterior convection reaches down into the PNS region in the 3D (but not the 2D) simulation. This may explain the slightly increased neutrino luminosities and shock radii in the 3D simulation seen in Fig.\,\ref{fig:lum} and Fig.\,\ref{fig:pns} at these earlier times. Lastly, we see a turbulent ``teardrop" in the 3D simulation extending from $\sim$20 to $\sim$80 km in the first $\sim$40 ms after bounce, trailing off to both the inner and outer convective regions by $\sim$60 ms after bounce. By comparison, this feature is much smaller in extent and delayed to $\sim$40 ms after bounce in the 2D model. The PNS convective zone has a characteristic size of $\sim$10 km, a turnover time of $\leq$10 ms, and convective velocities of $\sim$1000 km s$^{-1}$. This is a manifestation of the stronger turbulence within 100 km at early times in the 3D simulation.

\begin{figure*}
\includegraphics[width=0.49\textwidth]{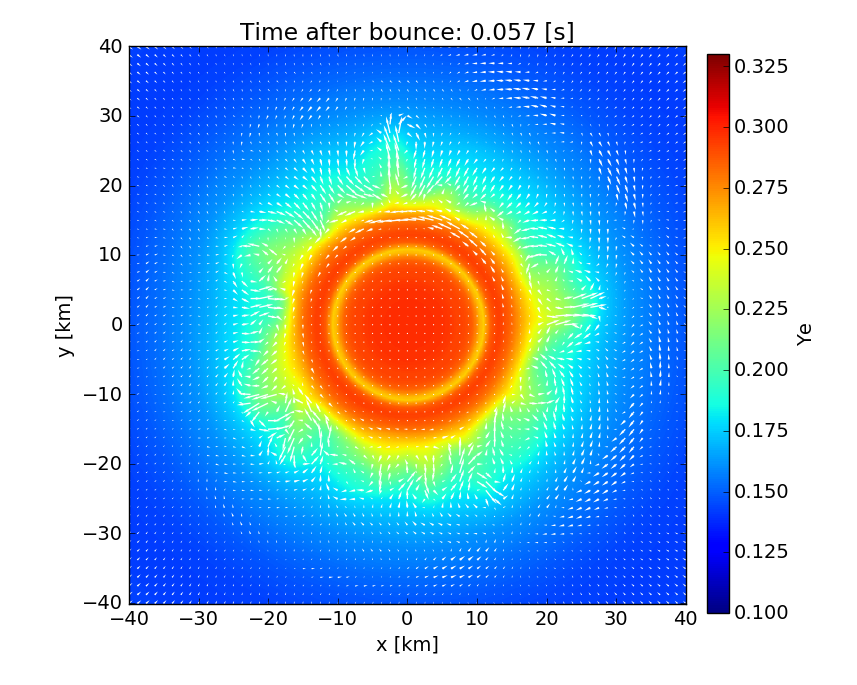}
\hfill
\includegraphics[width=0.49\textwidth]{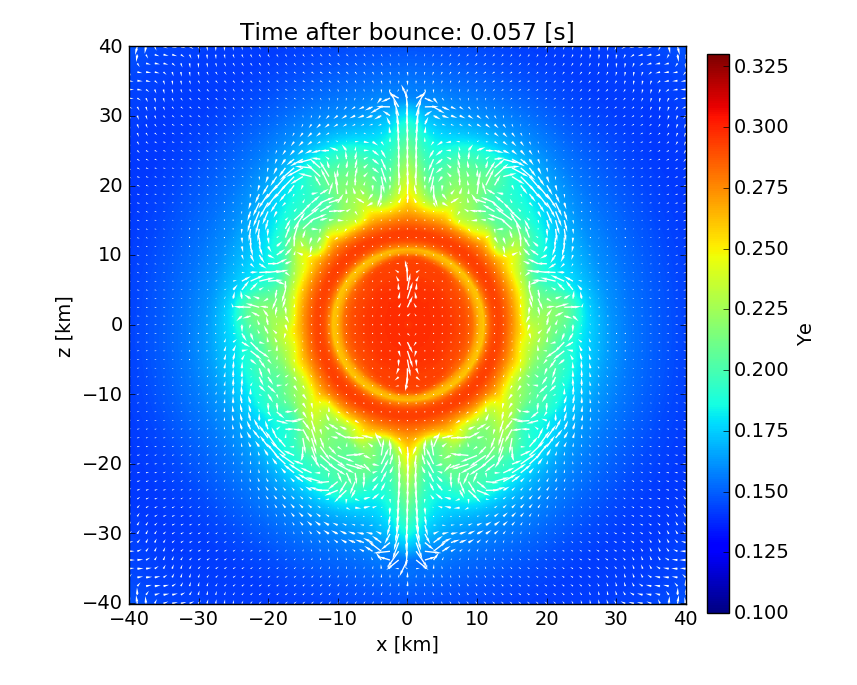}
\includegraphics[width=0.49\textwidth]{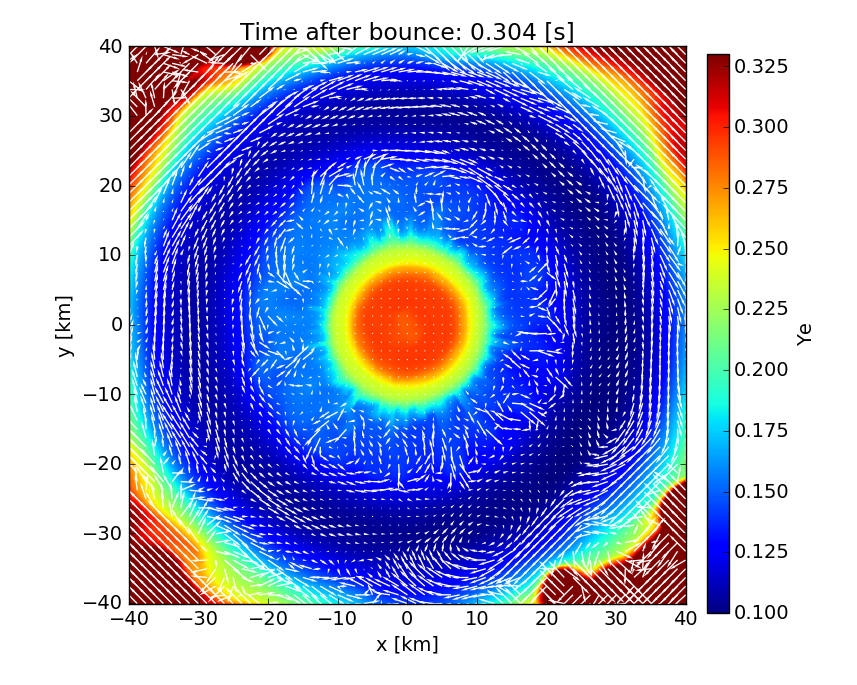}
\hfill
\includegraphics[width=0.49\textwidth]{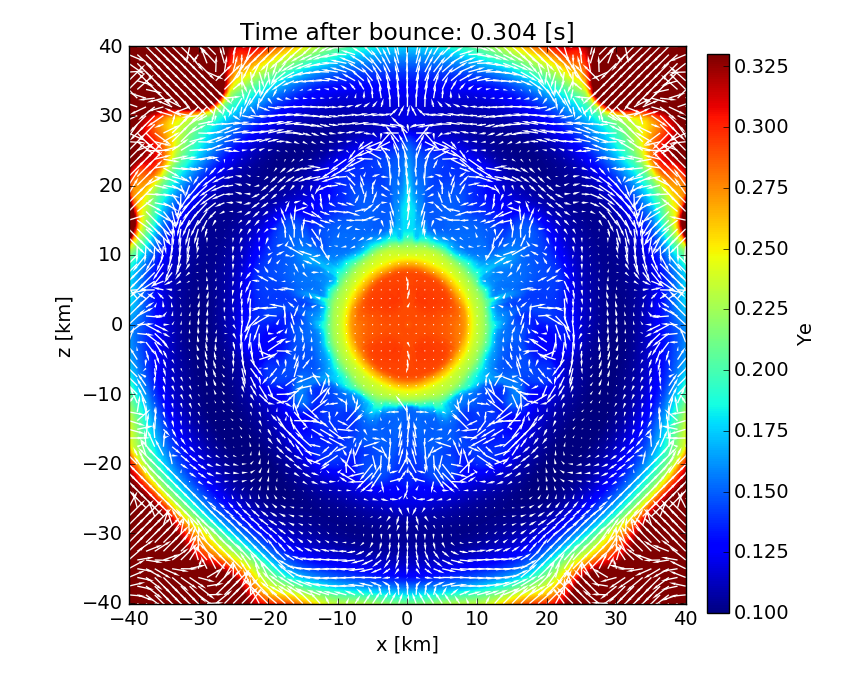}
\includegraphics[width=0.49\textwidth]{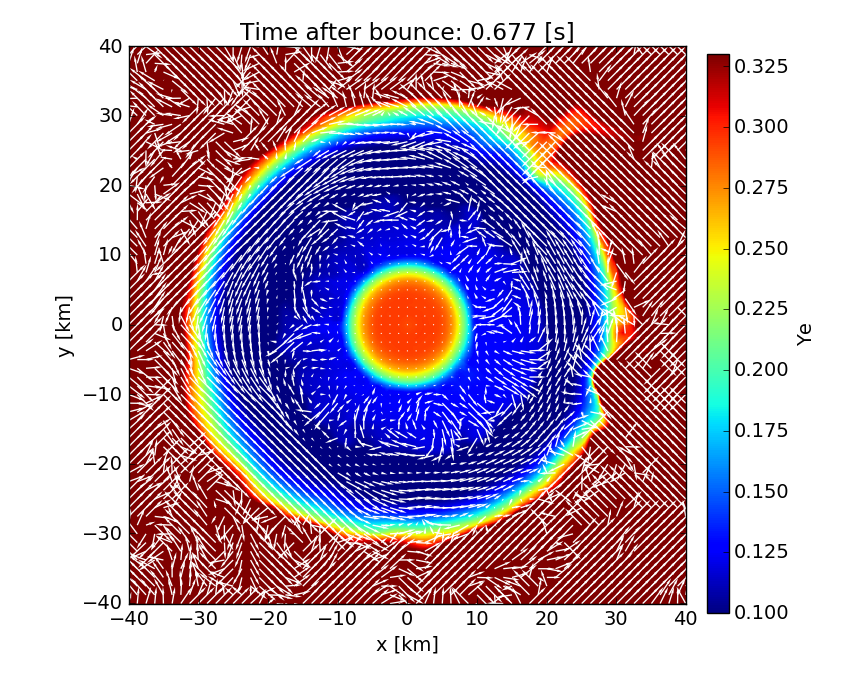}
\hfill
\includegraphics[width=0.49\textwidth]{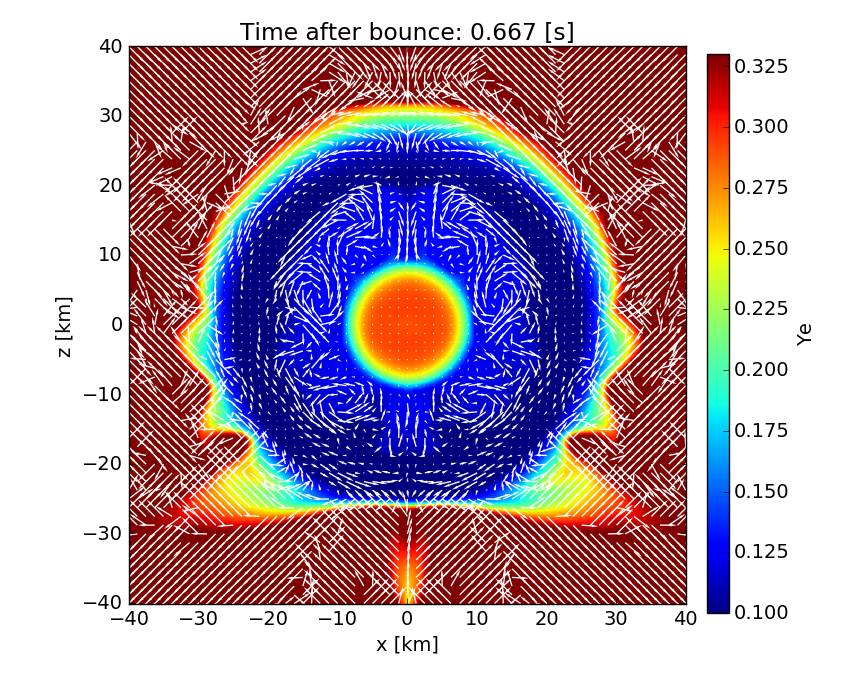}
\caption{Velocity vectors (white) on a Y$_e$ colormap depicted on an x-y slice of the 3D simulation 
(\textbf{left}) and an x-z slice of the 2D simulation (\textbf{right}) at $\sim$57 (\textbf{top}), 
$\sim$304 (\textbf{middle}), and $\sim$667 (\textbf{bottom}) ms after bounce to illustrate 
the evolution of inner-PNS convection. The velocity vector lengths are scaled to velocity and 
saturate at 2000 km s$^{-1}$. Note the characteristic convective whorls forming within the first $\sim$60 ms 
after bounce. The region of inner convection (with Y$_e$ $\sim$0.15-0.2) shrinks with the PNS, and 
at later times the exterior, neutrino-driven convective region (with Y$_e$ $\gtrapprox$0.3) is 
visible beyond $\sim$30 km, with low-Y$_e$ ``flares" traversing the boundary.}
\label{fig:ye_slide}
\end{figure*}

We explore the convective differences in the 2D and 3D simulations in Fig.\,\ref{fig:ye_slide}. We show velocity vectors (white) on a Y$_e$ colormap depicted on an x-y slice of the 3D simulation (left) and an x-z slice of the 2D simulation (right) at $\sim$57 (top), $\sim$304 (middle), and $\sim$667 (bottom) ms after bounce to illustrate the evolution of inner PNS convection. The vectors lengths are scaled to velocity and made to saturate at 2000 km s$^{-1}$. Note the characteristic convective whorls forming within the first $\sim$60 ms after bounce. 

\subsection{On the Possible Presence of the LESA and the SASI}\label{sec:sasi}

The lepton-number emission self-sustained asymmetry (LESA) was proposed in \cite{2014ApJ...792...96T} as a neutrino-hydrodynamical instability resulting in $\nu_e - \bar{\nu}_{e}$ asymmetry. In an earlier work (\citealt{vartanyan2018a}), we  explored the possibility of LESA by examining the dipole harmonic component, a$_{10}$, of the net lepton number flux. In that paper, we concluded that, at least in 2D, the effect was negligible and speculated that the inference of LESA may be a consequence of the use of the ray-by-ray approximation to multi-dimensional neutrino transport.

\begin{figure}
\includegraphics[width=0.5\textwidth]{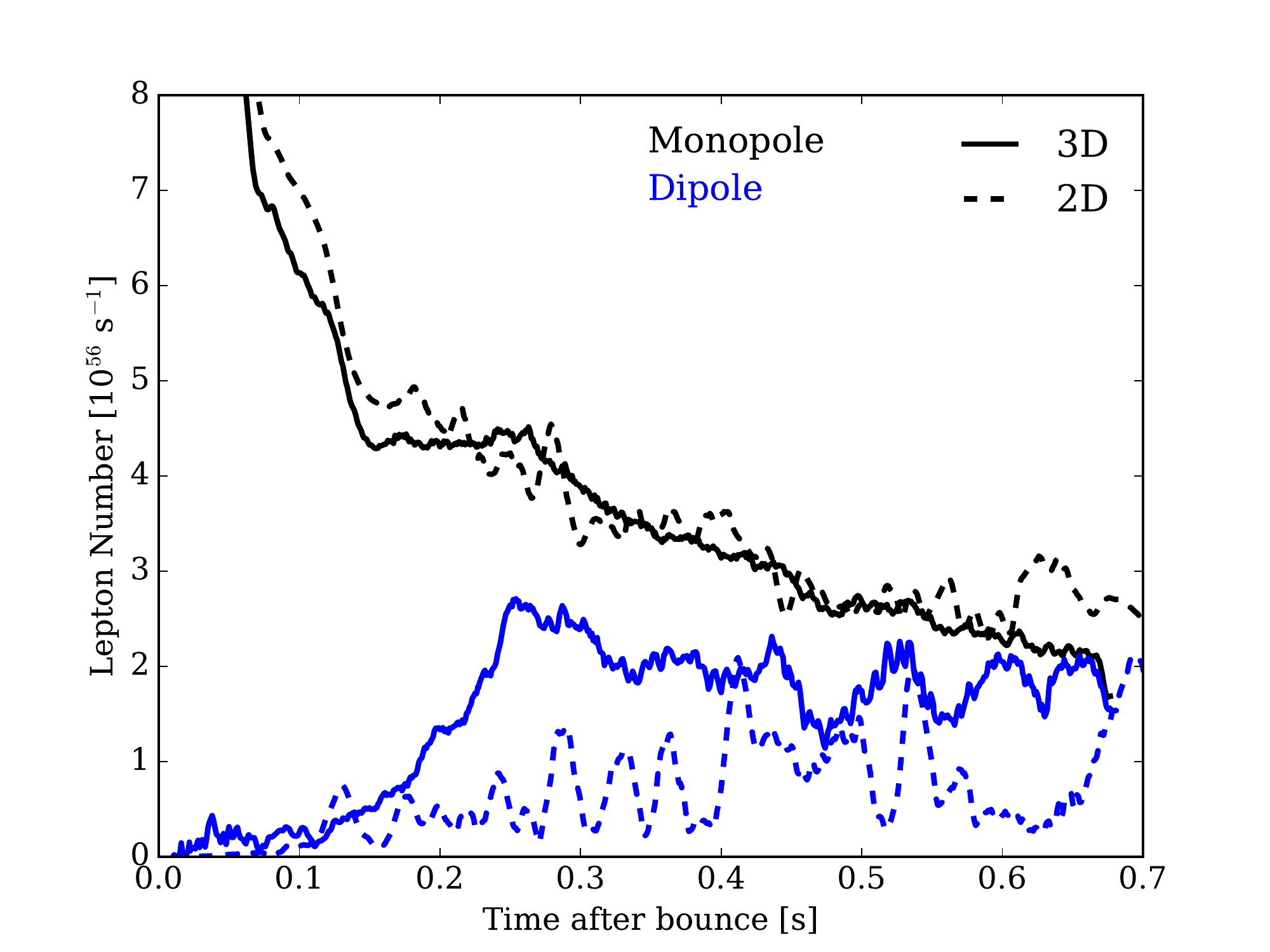}
\caption{We plot the monopole (black) and dipole (blue) of net lepton number asymmetry F$_{\nu_e}$ $-$ F$_{\bar{\nu}_{e}}$ (in units of 10$^{56}$ s$^{-1}$) as a function of 
time after bounce (in seconds) at 500 km to explore the possible appearance of the ``LESA" phenomenon. Solid 
indicates the 3D model and dashed the 2D model. We do see the LESA effect, and the dipole term in the 3D simulation is larger 
and less variable than in the corresponding 2D model. However, the dipole term becomes comparable in magnitude
to the monopole term only after $\sim$650 ms.}
\label{fig:lesa}
\end{figure}

We now extend our exploration of the possible presence of the LESA, using for the first time an exploding 3D model with full physical realism. In Fig.\,\ref{fig:lesa}, we depict the monopole and dipole components of the lepton asymmetry (F$_{\nu_e}$ $-$ F$_{\bar{\nu}_{e}}$) as a function of time after bounce at 500 km for both our 3D and 2D simulations. Here, we follow \cite{oconnor_couch2018} and plot instead the dipole magnitude,
\begin{equation}
A_{\mathrm{dipole}} = 3 \times \sqrt{\sum_{i=-1}^{1} a_{1i}^2},,
\end{equation}
using the normalization scheme of \cite{burrows2012}. The net effect is to increase the strength of the dipole term relative to the monopole term by a factor of $\sim$1.73 (3/$\sqrt{3}$).  We conclude that we do indeed find a LESA (see also \citealt{oconnor_couch2018}) effect, and that (at least for these models) it is stronger in 3D than in 2D.  However, the magnitude of the fluctuations in the lepton asymmetry is larger in 2D than in 3D.  In addition, whereas \cite{2014ApJ...792...96T} find that the dipole term overtakes the monopole term as early as $\sim$200 ms after bounce, we find that only after $\sim$650 ms after bounce does the dipole component of the LESA become comparable to the monopole term. We continue to suggest that the ray-by-ray approach leads to a larger LESA, but this remains to be tested with a comparison of 3D ray-by-ray and multi-angle simulations.

\begin{figure}
\includegraphics[width=0.5\textwidth]{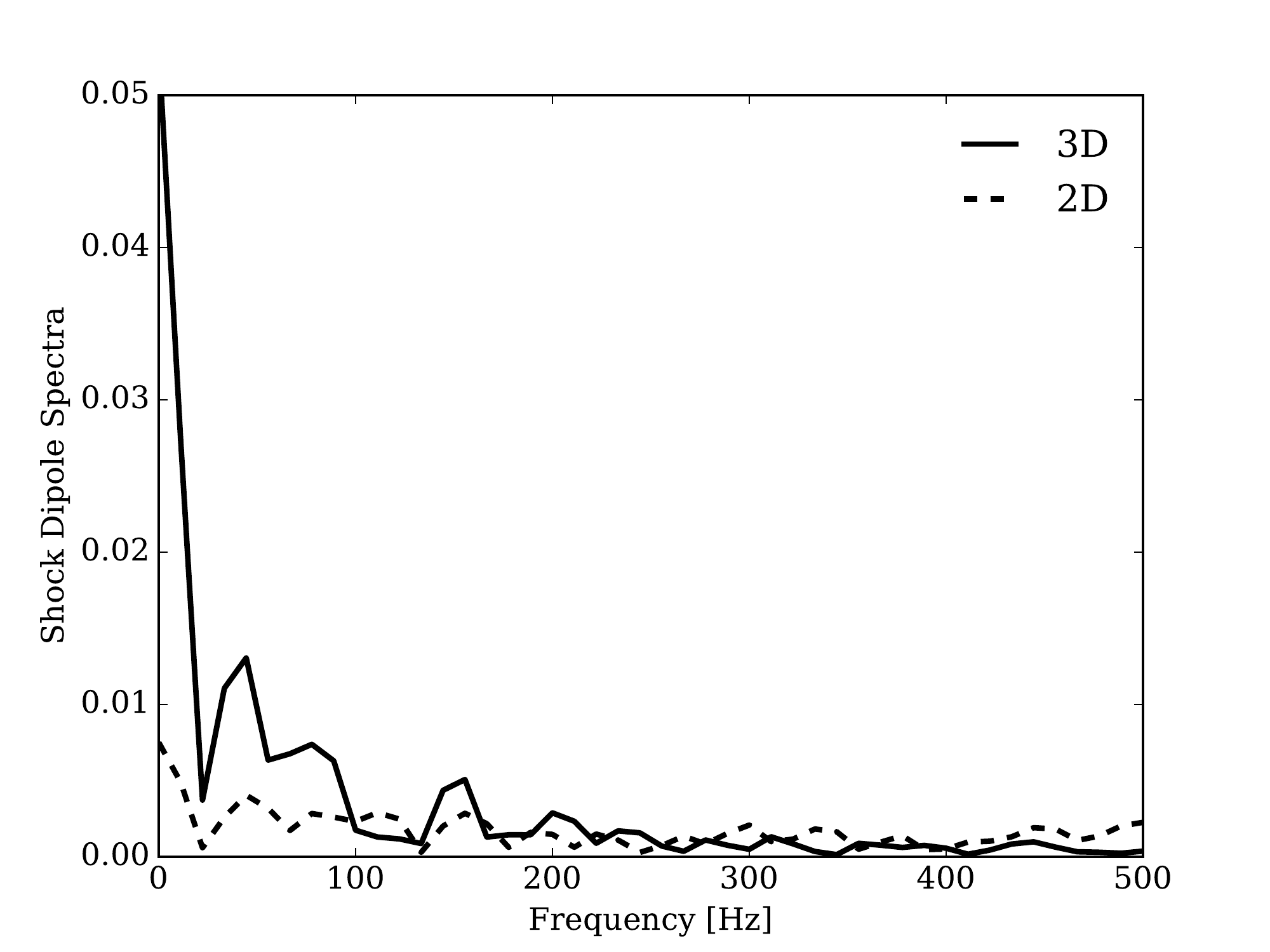}
\caption{We plot the Fourier decomposition of the shock radius dipole component as a function of frequency (in Hz) 
for the first 100 ms after bounce for the 3D (dashed) and 2D (solid) simulations. Note that while the 
dipole component is insignificant for both models early on, it is larger for the 3D model during the first $\sim$100 ms. This is also as seen in  Fig.\,\ref{fig:3} (solid red line, right panel).}
\label{fig:sasi}
\end{figure}

We have also studied our 3D simulation for the possible presence of the standing accretion shock instability (SASI) (\citealt{blondin2003}) during any phase of its evolution. If present, this should manifest in a narrow and obvious frequency peak in various power spectra. Recent work in 3D (\citealt{walk2018}) found pronounced peaks in the electron anti-neutrino power spectrum at $\sim$60 and $\sim$110 Hz that the authors associated with the SASI. In addition, \citealt{2017ApJ...851...62K} and \citealt{2016ApJ...829L..14K} suggested that softer equations of state manifest the SASI, with its gravitational wave signature lasting for an interval of $\sim$100 ms (a fraction of their simulation time) at frequencies of $\sim$50-200 Hz. Figure\,\ref{fig:sasi} portrays the Fourier decomposition of the dipole moment of the shock radius in both our 3D simulation and the associated 2D simulation out to 200 ms after bounce.  We find no clear peak at these frequencies, either by this metric or in the gravitational wave emissions (not shown here). Moreover, a glance at Fig.\,\ref{fig:3} reiterates that we see in the 3D run no significant dipole term in the shock radius until after explosion. \footnote{However, the dipole term is slighter stronger in the first $\sim$100 ms for the 3D simulation than for its 2D counterpart.} Therefore, we conclude that we have no evidence for the SASI in our 16-M$_{\odot}$ simulations. However, since we find an early explosion in both the 3D and 2D simulations, perhaps the SASI may have had insufficient time to develop. It is important to note, however, that \cite{oconnor_couch2018} likewise did not see a SASI for their 3D simulation (which was carried out to $\sim$600 ms after bounce, and did not explode) of a 20-M$_{\odot}$ progenitor when incorporating velocity dependence. Note that the small bump at $\sim$40 Hz in Fig.\,\ref{fig:3} corresponds to small-amplitude oscillations of the shock dipole in the first $\sim$100 ms after bounce. This feature is easily associated with the characteristic large-scale advective and convective time scales in the region between the shock and the PNS. 

We summarize here some of the catalysts to explosion in 3D. The 
16-M$_{\odot}$ progenitor model (\citealt{wh07}) upon which we focus in 
this paper has a steep density dropoff interior to 1.7-M$_{\odot}$ due to 
its Si-O interface. Such a sharp density drop has been shown to faciliate 
explosion in models incorporating turbulence (be they 2D or 3D) 
(\citealt{vartanyan2018a}) and we witnessed the explosion of this model in 
our previous 2D study. In addition, our inclusion of the many-body effect 
on the neutrino-nucleon scattering cross sections (\citealt{burrows2018}) 
and the introduction of the significant velocity perturbations to the 
progenitor are both conducive to explosion. These aspects, in addition to 
the effects of GR and heating due to inelastic neutrino-electron and 
neutrino-nucleon scattering, seem to be some of the agents of ``success."

\section{Conclusions}\label{sec:conc}

  We have presented in this paper one of the first non-rotating, state-of-the-art, 
full-microphysics simulations in three spatial dimensions to explode
as a supernova.  The explosion of a 16-M$_{\odot}$ progenitor is fully 
underway by $\sim$200 ms after bounce and at the end of the 
simulation is accumulating energy at a rate that if continued would reach
$\sim$0.5 Bethes ($0.5\times 10^{51}$ erg) within two seconds.  However, what
its final asymptotic energy will be remains to be seen.  The gravitational mass of the
remaining neutron star is $\sim$1.42 M$_{\odot}$.  The morphology of the
emerging debris field has a roughly dipolar structure, with two asymmetric
wide-angle lobes (one large, one small), whose axis emerged randomly.  Whether 
slight rotation would impose an axis for the ejecta, or what rotation rate would be necessary to
bias the emergent explosion axis, is not here determined.  By the end of the simulation,
an exploding debris field is accompanied by simultaneous inward accretion between 
the expanding lobes of some of the inner-progenitor matter, partly 
responsible for maintaining a driving neutrino luminosity (\citealt{burrows2007_features}). Interestingly,
the majority of the ejecta of this supernova are proton-rich, with Y$_e$ between 0.5 and 0.56.
This will have interesting consequences for the associated nucleosynthesis,
with the potential to explain in part the first r-process peak and p-process yields 
(\citealt{hoffman1996,pruet2006,frohlich2006,wanajo2011,frebel2018,bliss2018}). 

It has been shown in the past that vigorous turbulent convection behind 
the temporarily stalled shock is essential to ignite an explosion for almost
all anticipated progenitor structures. Only the rare progenitors at the lowest ZAMS masses
with very steep density profiles exterior to the collapsing Chandrasekhar core
explode in spherical symmetry (\citealt{kitaura2006,burrows2007_onemg,radice2017a}). 
The turbulent motions, boasting as they do a large effective
`$\gamma$' connecting kinetic energy with pressure/stress, are one agency.  Another is
the consequently larger gain region in the multi-D turbulent context.  A third could
be the longer dwell times in the gain region occasioned by the non-radial motions 
(\citealt{2008ApJ...688.1159M}). Aside from the necessity in most cases of the turbulence 
enabled in the multi-D context, the specific progenitor density profile is 
a major determinant, though the dependence upon the associated ``compactness" 
parameter (\citealt{2013ApJ...762..126O}) of the ``explodability" of a model is 
non-monotonic in subtle ways (Burrows et al. 2018).  Models with the 
lowest compactness may explode even in 1D via a wind mechanism (\citealt{burrows1987_wind}). However,
models with slightly higher compactness have trouble exploding (\citealt{burrows2018,2017IAUS..331..107O}), 
while models with even higher compactness (such as the 16-M$_{\odot}$ 
of this paper) explode rather easily.  Clearly, the explodability's dependence upon 
progenitor density profile is not straightforward.  

One aspect of this nuanced behavior is the role of the accretion through the shock 
of the silicon-oxygen interface (see \citealt{vartanyan2018a,2018ApJ...852...28S}).  The jump up 
in entropy at that interface is accompanied by a corresponding drop in mass density.  
If that drop is large and sharp, then when that interface is accreted through the stalled shock 
the confining ram pressure temporarily and abruptly declines, while not immediately 
altering the driving neutrino luminosities (emanating from the core) and heating rates.  
The consequence is often (as in the case studied in this paper) a kick into explosion, 
which in the immediate term is generally irreversible due to the quick diminution 
of neutrino cooling occasioned by expansion and the maintenance of heating.  However,
the magnitude and radius of this interface and the overall density profile
of the core at collapse are functions of stellar evolution (and stellar 
progenitor models), emphasizing the centrality to the viability and character of
core-collapse supernova explosion phenomenology of these initial states.

Another progenitor determinant of explosion may be its initial seed perturbations.  
It has been shown (\citealt{2013ApJ...778L...7C,2015ApJ...799....5C,muller_janka_pert,muller2016,burrows2018}) 
that if the seeds are of sufficient strength, then the ability of turbulence to ignite explosion 
is enhanced.  In the simulation highlighted in this paper, we imposed a modest physical
perturbation to the accreted velocity field that may have helped or accelerated explosion.  
However, whether perturbations are important, or merely facilitators, has not been determined
and the next generation of fully-3D progenitor models may illuminate this question 
(\citealt{2015ApJ...808L..21C,2016ApJ...833..124M}).

Those realistic physical processes that were conducive to the 3D explosion we witnessed in
this paper include neutrino-driven turbulence (\citealt{bhf1995,herant1994}), 
the net effects of general relativity (\citealt{bruenn_rel}), the inclusion 
of inelastic scattering and energy redistribution via neutrino-electron and neutrino-nucleon 
scattering (\citealt{burrows2018,vartanyan2018a,just2018}), the 
many-body correction to neutrino-nucleon scattering 
(\citealt{1998PhRvC..58..554B,PhysRevC.95.025801,burrows2018}), the accretion of a sharp 
silicon-oxygen interface at a propitious time (\citealt{vartanyan2018a}), and the imposition 
of velocity perturbations in the progenitor.
A major consequence of the many-body correction is the decrease in the scattering rate that
increases the neutrino emission rates.  This is particularly true for the $\nu_{\mu}$s,
and the resulting acceleration of core contraction leads to, among other things, the increase 
in the temperatures around the $\nu_e$ and $\bar{\nu}_e$ neutrinospheres.  This leads to a 
slight hardening of the emergent $\nu_e$ and $\bar{\nu}_e$ spectra and an increase
in the heating rate due to charged-current absorption on the free nucleons in the gain region.
One of the most important future classes of investigations of direct relevance to the
CCSN mechanism is the magnitude and role of many-body corrections to both the neutral-current
and the charged-current (\citealt{sawyer1999,roberts2012,roberts_reddy2017})
neutrino-matter interaction rates.  We note as well that even though the number of viable 
published nuclear equations of state is dwindling, the EOS dependence of the outcome of collapse has 
not been definitively addressed, nor well explained.  This will be a necessity in the years 
to come as laboratory constraints become ever more stringent. 

While the results presented in this paper are quite encouraging, there remain a number of 
important caveats.  Important among these are the dependence upon the spatial and energy-group 
resolutions.  In 3D, a resolution study, even with modern codes such as F{\sc{ornax}}, is 
expensive, but will be necessary to determine both the quantitative and qualitative 
limitations of what we have presented here.  The chaotic character of turbulent flow will make this a challenging
endeavor for the community going forward. Moreover, we have conducted these calculations including 
the effects of general relativity in approximate fashion (\S\ref{redshift}). Doing these calculations
with full GR will be important and attempts in this direction have already been made 
(\citealt{2016ApJ...831...98R,ott2018_rel,2018MNRAS.477L..80K}).  
To enable these forefront simulations, we still had to make 
approximations in the neutrino sector.  Foremost among these is 
the use of the moment formalism and an analytic closure for the second and third moments.  
While recent tests of the accuracy of such an approach in the core-collapse context 
are encouraging (\citealt{richers_nagakura2017,oconnor_code2018} ), solving the full Boltzmann equation with neutrino 
angles in the full six-dimensional phase space will require a significantly more capable 
national and international computational infrastructure.  Finally, it has been shown that 
explodability when near criticality and in multi-D is a sensitive function of details 
in the neutrino-matter interaction rates (\citealt{burrows2018}) in a way not seen in 1D 
simulations. This puts a premium on implementing correctly the correct microphysics.
All modelers aspire to this goal, but whether we or others actually have achieved this 
is, or should be, a constant worry. 

The model we presented was non-rotating.  We think that most collapsing cores, while 
they are certainly rotating, are not generically rotating at rates sufficient to 
make a qualitative difference most of the time 
(\citealt{emmering1989,faucher_kaspi,popov2012,noutsos2013}). However, this remains
to be exhaustively explored. Rapid rotation can certainly effect the outcome, 
both directly and by providing significant free energy to feed large magnetic 
fields and enable the direct effects of magnetic stress, when strong, on the explosion 
dynamics (see, e.g., \citealt{burrows2007_mag,mosta2015}). In fact, rapid rotation alone 
can affect the dynamics and facilitate explosion even when the expected magnetic field amplifications
are ignored (\citealt{2002ApJ...574L..65F,2004ApJ...601..391F,marek_janka2009,2018ApJ...852...28S}). Moreover, rapid 
rotation can also generate a non-axisymmetric spiral-arm mode, which resembles the 
SASI in the rotating context and might enlarge the gain region 
and, thereby, facilitate explosion (\citealt{takiwaki2016,2018ApJ...852...28S}).  Curiously, 
if the explosion is suitably delayed, such a mode may also grow in the 
non-rotating context (\citealt{blondin_shaw,rantsiou,guilet_fernandez,oconnor_couch2018}). 
This and other related issues are fruitful topics for future work.

However, we view the achievement of a 3D simulation that leads naturally to explosion, 
with competitive resolution, including all the relevant microphysics, using a 
state-of-the art simulation tool, and calculating significantly post-bounce as a major
milestone in the decades-long quest to resolve the core-collapse supernova puzzle in 
quantitative detail. What remains in the near term is to determine the progenitor mass dependence
of the outcome of collapse in 3D, to understand the possible roles of rotation, to explain 
the supernova energies and neutron star masses observed, and to explain the morphologies 
of the debris fields seen in supernova remnants.  Furthermore, a major motivation of all 
supernova simulations is the detailed explanation of the explosive production of the 
elements. The ejecta we find are mostly proton-rich, and this emerges naturally from the 
detailed simulations.  What the consequences are of this finding will be one
of the topics of our future studies as we continue our quest to understand
one of the most persistent problems in stellar and nuclear astrophysics.










\section*{Acknowledgments}

The authors acknowledge helpful discussions with Todd Thompson regarding inelastic scattering, Evan O'Connor regarding the equation of state, and Gabriel Mart\'inez-Pinedo concerning electron capture on heavy nuclei. We also acknowledge invaluable support from Viktoriya Morozova with visualization using VisIt, and Sydney Andrews for helpful discussion and feedback. We acknowledge support from the U.S. Department of Energy Office of Science and the Office of Advanced Scientific Computing Research via the Scientific Discovery through Advanced Computing (SciDAC4) program and Grant DE-SC0018297 (subaward 00009650). In addition, we gratefully acknowledge support from the U.S. NSF under Grants AST-1714267 and PHY-1144374 (the latter via the Max-Planck/Princeton Center (MPPC) for Plasma Physics). DR acknowledges partial support as a Frank and Peggy Taplin Fellow at the Institute for Advanced Study. The authors employed computational resources provided by the TIGRESS high performance computer center at Princeton University, which is jointly supported by the Princeton Institute for Computational Science and Engineering (PICSciE) and the Princeton University Office of Information Technology, and by the National Energy Research Scientific Computing Center (NERSC), which is supported by the Office of Science of the US Department of Energy (DOE) under contract DE-AC03-76SF00098. The authors express their gratitude to Ted Barnes of the DOE Office of Nuclear Physics for facilitating their use of NERSC. This overall research project is also part of the Blue Waters sustained-petascale computing project, which is supported by the National Science Foundation (awards OCI-0725070 and ACI-1238993) and the state of Illinois. Blue Waters is a joint effort of the University of Illinois at Urbana-Champaign and its National Center for Supercomputing Applications. This general project is also part of the ``Three-Dimensional Simulations of Core-Collapse Supernovae" PRAC allocation support by the National Science Foundation (under award \#OAC-1809073).  Under the local award \#TG-AST170045, our ongoing supernova efforts are enhanced through access to the resource Stampede2 in the Extreme Science and Engineering Discovery Environment (XSEDE), which is supported by National Science Foundation grant number ACI-1548562. This work was performed under the auspices of the U.S. Department of Energy by Lawrence Livermore National Laboratory under Contract DE-AC52-07NA27344 and has been assigned an LLNL document release number LLNL-JRNL-757405. This work was performed under the auspices of the U.S. Department of Energy by Los Alamos National Laboratory under Contract DE-AC52-06NA25396 and has been assigned an LANL document release number LA-UR-18-28730. JD acknowledges support from the Laboratory Directed Research and Development program at Los Alamos National Laboratory.

\bibliographystyle{mnras}
\bibliography{References}

\newpage

\appendix
\onecolumn

\section{Approximate General Relativistic Formulation}
\label{redshift}

As stated in Sec.\,\ref{sec:1}, we use the M1 closure to truncate the radiation moment hierarchy 
by specifying the second and third moments as algebraic functions in terms of the zeroth and first. The basic 
equations of radiative transfer in the comoving frame that we solve are the zeroth- and first-moment
equations of the full equation of radiative transfer for the specific intensity.
For neutrino transfer, we currently
follow the evolution of $\nu_e$, $\bar{\nu}_e$, and $\nu_x$ neutrinos, where the latter represents
the $\nu_{\mu}$, $\bar{\nu}_{\mu}$, and $\nu_{\tau}$, neutrinos collectively. With explicit neutrino source and sink terms, the 
full set of Newtonian radiation/hydrodynamic equations, with a specific focus on the neutrino radiation implementation
relevant to the study of core-collapse supernovae, are:

\begin{align}
\rho_{,t} + (\rho v^i)_{;i}  &= 0\, , \\
(\rho v_j)_{,t} + (\rho v^i v_j + P \delta^i_j)_{;i} &= -\rho \phi_{,j} + c^{-1} \sum_s \int_0^\infty (\kappa_{s\varepsilon} + \sigma^{\rm tr}_{s\varepsilon}) F_{s\varepsilon j} d\varepsilon\, , \\
\left[\rho \left(e + \frac{1}{2} \lVert v\rVert^2\right)\right]_{,t} + \left[\rho v^i \left(e + \frac{1}{2} \lVert v\rVert^2 + \frac{P}{\rho}\right)\right]_{;i} &= -\rho v^i \phi_{,i} - \sum_s \int_0^\infty \left(j_{s\varepsilon} - c \kappa_{s\varepsilon} E_{s\varepsilon} - \frac{v^i}{c} (\kappa_{s\varepsilon} + \sigma^{\rm tr}_{s\varepsilon}) F_{s\varepsilon i}\right) d\varepsilon\, , \\
(\rho Y_e)_{,t} + (\rho Y_e v^i)_{;i} &= \sum_s \int_0^\infty \xi_{s\varepsilon} (j_{s\varepsilon} - c \kappa_{s\varepsilon} E_{s\varepsilon}) d\varepsilon\, ,\\
E_{s\varepsilon,t} + (F_{s\varepsilon}^i + v^i E_{s\varepsilon})_{;i} - v^i_{;j}\frac{\partial}{\partial\ln\varepsilon} P_{s\varepsilon i}^j &= j_{s\varepsilon} - c \kappa_{s\varepsilon} E_{s\varepsilon}\, , \\
F_{s\varepsilon j,t} + (c^2 P_{s\varepsilon j}^i + v^i F_{s\varepsilon j})_{;i} + v^i_{;j} F_{s\varepsilon i} - v^i_{;k} \frac{\partial}{\partial\varepsilon} (\varepsilon Q^k_{s\varepsilon ji}) &= -c(\kappa_{s\varepsilon} + \sigma^{\rm tr}_{s\varepsilon}) F_{s\varepsilon j},\, 
\end{align}

where $e$ is the specific internal energy, $P=P(\rho,e,Y_e)$ is the pressure, $\rho$ is the mass density,
$Y_e$ is the electron fraction, $v_i$ are the velocity components, $\kappa_{s\varepsilon}$ and
$\sigma^{\rm tr}_{s\varepsilon}$ are the absorption and transport scattering
opacities, and $s\in\{\nu_e,\bar{\nu}_e,\nu_x\}$, where
\begin{equation}
\xi_{s\varepsilon} = \begin{cases}
        -(N_A \varepsilon)^{-1}&        \text{$s=\nu_e$},\\
        (N_A \varepsilon)^{-1}&         \text{$s=\bar{\nu}_e$},\\
        0&                                                      \text{$s=\nu_x$}\, .
        \end{cases}
\end{equation}

The corresponding radiation energy and momentum equations modified to approximately incorporate general relativistic effects are:
\begin{align}
E_{s\varepsilon,t} + (\alpha F_{s\varepsilon}^i + {\bf v^i E_{s\varepsilon}})_{;i} - \alpha {\bf v^i_{;j}\frac{\partial}{\partial\ln\varepsilon} P_{s\varepsilon i}^j} &= \alpha (j_{s\varepsilon} - c \kappa_{s\varepsilon} E_{s\varepsilon}) + \alpha G^e\, ,\\
F_{s\varepsilon j,t} + (c^2 \alpha P_{s\varepsilon j}^i + {\bf v^i F_{s\varepsilon j}})_{;i} + \alpha {\bf v^i_{;j} F_{s\varepsilon i} - \alpha v^i_{;k} \frac{\partial}{\partial\varepsilon} (\varepsilon Q^k_{s\varepsilon ji})} &= -c \alpha (\kappa_{s\varepsilon} + \sigma^{\rm tr}_{s\varepsilon}) F_{s\varepsilon j} + \alpha G^m_j\, ,
\end{align}

where differentiation is indicated with standard notation, $\varepsilon$ is the neutrino energy, $s\in\{\nu_e,\bar{\nu}_e,\nu_x\}$,
$E_{s\varepsilon}$ is the radiation energy density spectrum (zeroth moment), $F_{s\varepsilon j}$ is radiation flux
spectrum (first moment),  $P_{s\varepsilon i}^j$ is the radiation pressure tensor (second moment), $Q^k_{s\varepsilon ji}$
is the heat tensor (third moment), $\alpha = \exp(\phi/c^2)$, and the other variables have their standard meanings.
$\phi$ is the gravitational potential. 

$G^e$ and $G^m_j$ are the main terms to add in order to include gravitational redshifts
(\citealt{rampp_janka2002,shibata2011}).  They are given by

\begin{align*}
G^e &= -{\bf F_{s\varepsilon}}\cdot{\bf \nabla}\phi/c^2 + {\bf \nabla}\phi/c^2\cdot\partial{(\varepsilon{\bf F_{s\varepsilon}})}/\partial\varepsilon\, ,\\
G^m_{\bf j} &= -E_{s\varepsilon}\nabla_{\bf j}\phi+\nabla_{\bf i}\phi\cdot\partial{(\varepsilon{\bf P^i_{s\varepsilon{j}}})}/\partial\varepsilon\, ,
\end{align*}

where $\nabla_{\bf j}\phi/c^2 = -g_j/c^2$, and one must do the contravariant/covariant raising
or lowering according to the metric.
Notice that the last terms in the equations for $G^e$ and $G^m_j$ integrate out to zero when one integrates over energy groups, leaving
the terms analogous to the ``$\rho {\bf v} \cdot {\bf g}$" and ``$\rho{\bf g}$" work and force terms.

The gravitational potential, $\phi$, is generally taken to be the ``GR-corrected" monopole term ($\phi_{\rm TOV}$).  There are a variety
of ways to approximate this, and the source for this approximation is \cite{marek2006}.  Many people use their ``Case A,"
as do we, though they say their ``Case B" is also very good.  The relevant equations are:

\begin{align}
\frac{d\phi_{\rm TOV}}{dr} &= G\frac{m_{\rm TOV} + 4\pi r^3(P + P_{\nu})/c^2}{r^2\Gamma^2}\left[\frac{\rho + E/c^2 + P/c^2}{\rho}\right]\, ,\\
\Gamma(r) &= \sqrt{1 + v^2/c^2 - \frac{2Gm_{\rm TOV}(r)}{rc^2}}\, ,\\
\frac{dm_{\rm TOV}}{dr} &= 4\pi r^2\left(\rho + E/c^2 + E_{\nu}/c^2 + \frac{{\bf v}\cdot {\bf F_{\nu}}/c^2}{\Gamma}\right)\Gamma\,\,\,\, ({\bf Case\ A})\, ,\\
\frac{dm_{\rm TOV}}{dr} &= 4\pi r^2 \rho \,\,\,\,  ({\bf Case\ B}) \, ,
\end{align}
where $\rho$ is the rest mass density, $P_{\nu}$ is the total neutrino pressure, $E$ is the total matter internal energy density,
$P$ is the matter pressure, $v$ is some averaged radial speed, $E_{\nu}$ is the total neutrino energy density, and $F_{\nu}$
is the total neutrino flux.  If one is using a multipole expansion to derive the potential, then the potential used is:
\begin{equation}
\phi_{\rm eff} = \phi - \bar{\phi} + \phi_{\rm TOV}\, ,
\end{equation}
where $\phi$ is the multipole Newtonian potential, $\bar{\phi}$ is the monopole Newtonian potential, and $\phi_{\rm TOV}$
is the monopole TOV potential.
Note that this equation merely subtracts out the monopole term from the total
Newtonian potential, and then adds it back, corrected for the GR effects in the approximate way suggested by \cite{marek2006}.

So, one needs to calculate the total pressure, energy density, and flux of the neutrinos,
calculate $\langle v^2/c^2\rangle$ as a function of radius, and then integrate radially/spherically to get $m_{\rm TOV}$ and $\phi_{\rm TOV}$.
All the non-monopolar potential contributions (if included) are Newtonian; only the monopole term is adjusted approximately for GR.
This correction to the monopolar potential used in the matter momentum equation is likely the dominant effect of GR.  However,
the ``GR-corrected" transport scheme above incorporates the neutrino energy redshifts, as well as the time dilation.
In the steady-state, zero-motion limit, the total luminosity $\times$ exp(2$\phi/c^2$) is a constant, as it should be (note the factor of 2).
\section{Neutrino-Matter Interactions for the Supernova Problem}
\label{sec.nu_inter}

A comprehensive set of neutrino-matter interactions is 
implemented into F{\sc{ornax}}, and these are described in 
\cite{2006NuPhA.777..356B}. They include neutrino-nucleon scattering and
absorption and neutrino-nucleus scattering with ion-ion-correlations, 
weak screening, and form-factor corrections.
For the neutrino-nucleon scattering cross sections, we use eq. 24 of
\cite{2006NuPhA.777..356B} and for the charged-current absorption cross sections
(i.e., $\nu _e + n \rightarrow p + e^-$ and $\bar{\nu}_e + p \rightarrow n + e^+$)
we follow the approach described in sections 3.1 and 3.2 of \cite{2006NuPhA.777..356B}.
Emissivities are calculated to be consistent with Kirchhoff's law of detailed
balance, as described with eqs. 7 and 8 of \cite{2006NuPhA.777..356B}.
The latter incorporates the stimulated emission term for fermions, which
in other formulations (cf. \citealt{1985ApJS...58..771B}) is equivalent to Pauli blocking.

The weak magnetism and recoil corrections for both absorption and
scattering are incorporated as multiplicative factors, using the linear
fits in neutrino energy of \cite{horowitz2002}. The linear fits are accurate
below $\sim$100 MeV for $\nu_e$s and $\sim$50 MeV for $\bar{\nu}_e$s, but
deviate at higher neutrino energies.  Nevertheless, higher-energy $\nu_e$s
reside only in the core.  Moreover, there are effectively no higher-energy
$\bar{\nu}_e$s in the supernova core. The degeneracy of $\nu_e$s that
results from lepton trapping also results in a positive $\nu_e$ chemical
potential, and, hence, a negative $\bar{\nu}_e$ chemical potential.  Such
a negative potential exponentially suppresses $\bar{\nu}_e$s. Where the
$\bar{\nu}_e$s can finally achieve significant occupancy, the temperatures
and densities are too low to produce many $\bar{\nu}_e$s above $\sim$60
MeV.  The $\nu_{\mu}$s don't experience such Boltzmann suppression, but are
thermally produced, and, hence, given the core temperatures, are not
produced in significant numbers above $\sim$100 MeV.  Be that as it may,
we have also implemented the fully non-linear weak-magnetism and recoil
formalism into F{\sc{ornax}}, done side-by-side comparison simulations in 1D,
and found that the hydrodynamic and radiation results are almost
identical. For the ``$\nu_{\mu}$" neutrino-nucleon scattering, we use the
average of the weak magnetism and recoil corrections for the neutrino and
anti-neutrino types.

Inelastic neutrino-nucleon scattering is handled using a modified version
of the \cite{2003ApJ...592..434T} approach. A comprehensive
discussion of the equations and techniques we employ to handle
inelasticity in both neutrino-nucleon and neutrino-electron scattering is found in section 4 of \cite{burrows_thompson2004}. The integrals for the
dynamic structure functions for inelastic neutrino-electron scattering are
handled relativistically, as in \cite{reddy1999} and the equations on pages
144-148 of \cite{burrows_thompson2004}. As in the case of inelastic
scattering off nucleons, all integrals for inelastic neutrino-electron
scattering are done numerically with quadratures, and the results are
tabulated in large tables for use during simulation when inelasticity and
redistribution are turned on (as they are in this work). This involves as
an intermediate step the numerical calculation of polylogarithmic
functions. Neutrino sources and sinks due to nucleon-nucleon
bremsstrahlung and electron-positron annihilation are included, as
described in \cite{thomp_bur_horvath}. We also include the
many-body correction to neutrino-nucleon scattering of \cite{PhysRevC.95.025801}, 
which is a variation of the formalism of \cite{1998PhRvC..58..554B}
extended to lower densities. This correction, which slightly decreases the
scattering rate at progressively higher densities, has been shown to
support the ``explodability" of supernova models and is physically
well-motivated.  A corresponding correction for charged-current absorption
onto nucleons has not been calculated in full (\citealt{roberts_reddy2017}), nor
included in F{\sc{ornax}}, but could be of equal relevance. For electron capture on nuclei, of importance during the infall phase, we employ the rates of \cite{2010NuPhA.848..454J}.


\appendix

\twocolumn

\begin{figure*}
\includegraphics[width=0.9\textwidth]{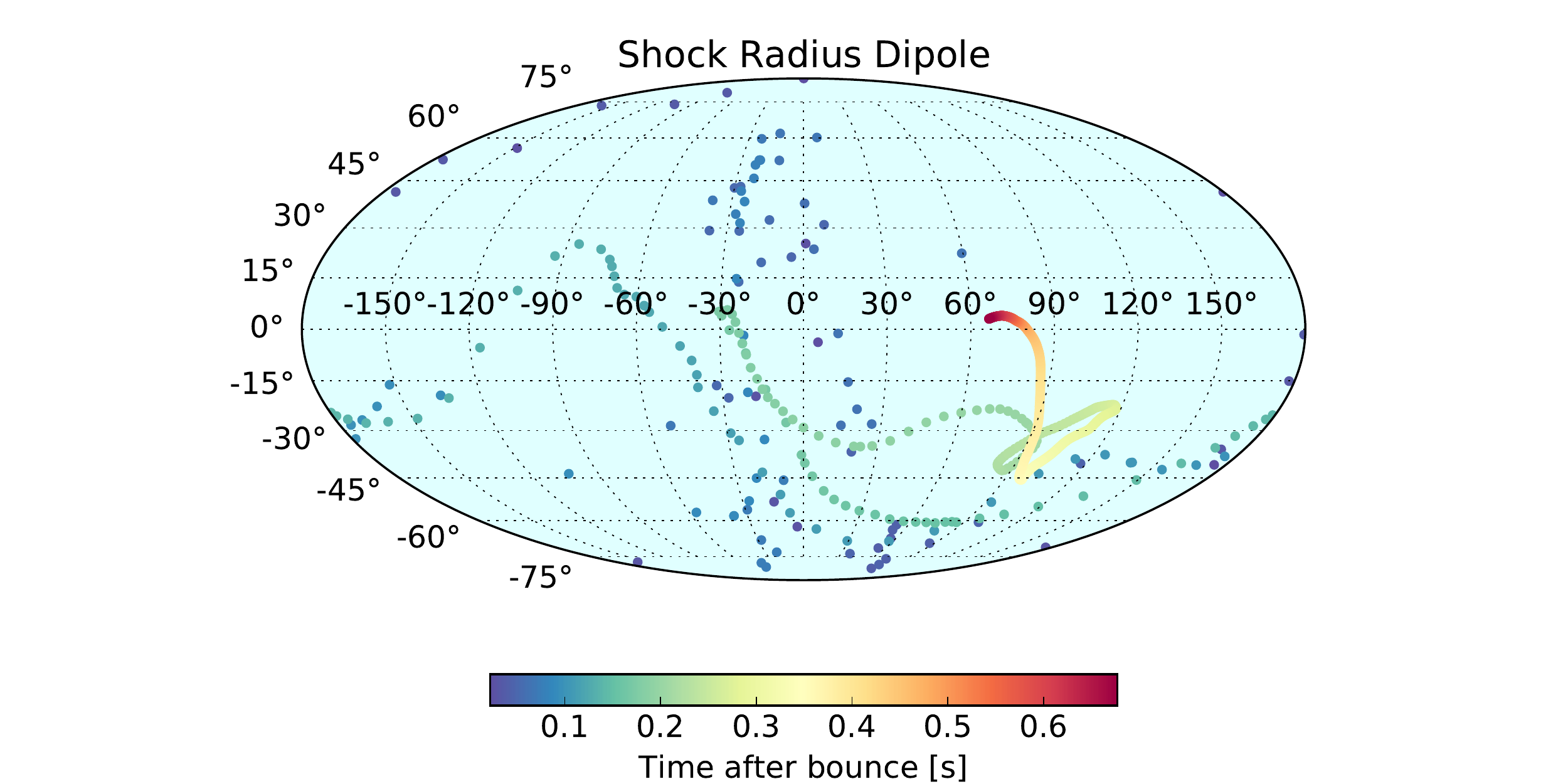}
\caption{A Mollweide projection of the direction of the shock dipole as a function of time (in seconds)
after bounce, color-coded. Early on, the shock dipolar direction is changes sporadically before settling at late times
to a randomly chosen axis. See Fig.\, 3 of \protect\cite{burrows2012} for a comparison. Note that, in a 2D simulation, the dipole axis is required to lie along the z-axis; this is not the case in a 3D simulation.}
\label{fig:4}
\end{figure*}

\label{lastpage}
\end{document}